\PassOptionsToPackage{hyphens}{url}
\PassOptionsToPackage{x11names,dvipsnames}{xcolor}
\documentclass[conference]{IEEEtran}

\usepackage{xcolor}
\usepackage{graphicx}
\usepackage{transparent}
\usepackage{tikz}
\usepackage{babel}
\usepackage[normalem]{ulem}
\usepackage{xurl}
\usepackage{verbatim}
\usepackage{hyperref}
\usepackage{listings}
\usepackage{varwidth}
\usepackage[super]{nth}
\usepackage{moreenum}
\usepackage{tikz}
\usepackage[absolute,overlay]{textpos}
\usepackage{tabfigures}
\usepackage{pgfplots}
\usepackage{siunitx}
\usepackage{textcomp}
\usepackage{multirow}
\usepackage{pgfcalendar}
\usepackage{tabularx}
\usepackage{etoolbox}
\usepackage{cleveref}
\usepackage{booktabs}
\usepackage{nicematrix}
\usepackage{cite}
\usepackage{makecell}

\pgfplotsset{compat=1.18}

\usepgfplotslibrary{units}
\usepgfplotslibrary{dateplot}
\usepgfplotslibrary{colormaps}
\usepgfplotslibrary{groupplots}

\pgfplotsset{colormap/viridis high res}

\pgfplotsset{
    /pgfplots/layers/standard with background/.define layer set={
        background,
        axis background,
        axis grid,
        axis ticks,
        axis lines,
        axis tick labels,
        pre main,
        main,
        axis descriptions,
        axis foreground
    }{/pgfplots/layers/standard}
}

\pgfmathdeclarefunction{lg10}{1}{%
    \pgfmathparse{ln(#1)/ln(10)}%
}

\usetikzlibrary{
  decorations.pathreplacing,
  positioning,
  shapes,
  calc,
  fit,
  tikzmark,
  backgrounds,
  arrows.meta,
  intersections
}

\tikzset{
  white background/.style={
    show background rectangle,
    tight background,
    background rectangle/.style={
      fill=white
    }
  }
}

\tikzstyle{global}=[
  white background,
]

\tikzstyle{txt}=[align=center, execute at end node=\vphantom{bg}]
\tikzstyle{tiny}=[node font=\tiny]
\tikzstyle{ss}=[node font=\scriptsize]
\tikzstyle{fn}=[node font=\footnotesize]
\tikzstyle{sm}=[node font=\small]
\tikzstyle{tiny txt}=[tiny, txt]
\tikzstyle{fn txt}=[fn, txt]
\tikzstyle{mas}=[midway, above, sloped]
\tikzstyle{mbs}=[midway, below, sloped]

\colorlet{good}{green!75!black}
\colorlet{bad}{red!75!black}
\colorlet{ok}{yellow!75!black}

\tikzstyle{good}=[green!75!black]
\tikzstyle{bad}=[red!75!black]
\tikzstyle{ok}=[yellow!75!black]

\tikzset{>={Classical TikZ Rightarrow[]}}
\tikzset{small >/.tip={Classical TikZ Rightarrow[scale=0.67]}}
\tikzset{smaller >/.tip={Classical TikZ Rightarrow[scale=0.5]}}

\definecolor{fbblue}{rgb}{.278, .404, .667}

\colorlet{samplebg}{gray!50}
\colorlet{sample1}{fbblue}
\colorlet{sample2}{gray}
\colorlet{sample3}{green!45!white!60!black}
\colorlet{sample4}{red!45!white!65!black}

\colorlet{graymain}{gray}
\colorlet{graylight}{graymain!70!white}
\colorlet{graylighter}{graymain!45!white}
\colorlet{graylightest}{graymain!25!white}
\colorlet{graydark}{graymain!70!black}
\colorlet{graydarker}{graymain!45!black}

\colorlet{bluemain}{fbblue}
\colorlet{bluelight}{bluemain!70!white}
\colorlet{bluelighter}{bluemain!45!white}
\colorlet{bluelightest}{bluemain!25!white}
\colorlet{bluedark}{bluemain!70!black}
\colorlet{bluedarker}{bluemain!45!black}

\colorlet{greenmain}{green!45!white!60!black}
\colorlet{greenlight}{greenmain!70!white}
\colorlet{greenlighter}{greenmain!45!white}
\colorlet{greenlightest}{greenmain!25!white}
\colorlet{greendark}{greenmain!70!black}
\colorlet{greendarker}{greenmain!45!black}

\colorlet{redmain}{red!45!white!65!black}
\colorlet{redlight}{redmain!70!white}
\colorlet{redlighter}{redmain!45!white}
\colorlet{redlightest}{redmain!25!white}
\colorlet{reddark}{redmain!70!black}
\colorlet{reddarker}{redmain!45!black}

\colorlet{yellowmain}{yellow!80!white!75!black}
\colorlet{yellowlight}{yellowmain!70!white}
\colorlet{yellowlighter}{yellowmain!45!white}
\colorlet{yellowlightest}{yellowmain!25!white}
\colorlet{yellowdark}{yellowmain!70!black}
\colorlet{yellowdarker}{yellowmain!45!black}

\colorlet{problemred}{red!15!white}
\colorlet{cautionorange}{orange!25!white}
\colorlet{warningyellow}{yellow!50!white}
\colorlet{highlightgreen}{green!15!white}
\colorlet{highlightblue}{fbblue!15!white}

\newlist{todolist}{itemize}{2}
\setlist[todolist]{label=$\square$,leftmargin=0.7cm}
\usepackage{pifont}

\newcommand{\fakeinterrobang}{{\ooalign{?\cr\kern.2ex!\cr}}}

\newcommand{\PreserveBackslash}[1]{\let\temp=\\#1\let\\=\temp}
\newcolumntype{C}[1]{>{\PreserveBackslash\centering}p{#1}}
\newcolumntype{R}[1]{>{\PreserveBackslash\raggedleft}p{#1}}
\newcolumntype{L}[1]{>{\PreserveBackslash\raggedright}p{#1}}

\setenumerate[0]{label={\tbfigures\arabic*.}}

\def\NAME{OpenZL}

\usepackage{amsmath,amssymb,amsfonts}
\usepackage{algorithm}
\usepackage{algpseudocode}
\usepackage{nicefrac}

\usepackage{amsthm}
\theoremstyle{definition}
\newtheorem{definition}{Definition}[section]

\newtheorem*{remark}{Remark}

\usepackage[T1]{fontenc}
\usepackage[scale=0.95]{sourcecodepro}

\newlist{enuminline}{enumerate*}{1}
\setlist[enuminline]{label=(\roman*)}

\colorlet{openzlplotcolor}{ForestGreen}
\colorlet{zstdplotcolor}{Cyan}
\colorlet{xzplotcolor}{Fuchsia}
\colorlet{zlibplotcolor}{Mahogany}
\colorlet{bloscplotcolor}{Magenta}
\colorlet{parquetplotcolor}{RoyalBlue}
\colorlet{cmixplotcolor}{CarnationPink}
\colorlet{nncpplotcolor}{Dandelion}

\tikzstyle{sao transform}=[draw, rounded corners=2pt, inner sep=4pt, font=\footnotesize]
\tikzstyle{sao stream node}=[fill=white, inner sep=1pt, font=\scriptsize]

\tikzstyle{rpbar}=[
  fill,
  draw=none,
]

\definecolor{metafg}{HTML}{1C2B33}
\definecolor{metabg}{HTML}{F1F4F7}

\pgfdeclareplotmark{openzlplotmark}{\pgfuseplotmark{*}}
\pgfdeclareplotmark{zstdplotmark}{\pgfuseplotmark{triangle*}}
\pgfdeclareplotmark{xzplotmark}{\pgfuseplotmark{diamond*}}
\pgfdeclareplotmark{zlibplotmark}{\pgfuseplotmark{star}}
\pgfdeclareplotmark{brotliplotmark}{\pgfuseplotmark{pentagon*}}
\pgfdeclareplotmark{bloscplotmark}{\pgfuseplotmark{+}}
\pgfdeclareplotmark{parquetplotmark}{\pgfuseplotmark{x}}

\newif\ifcommentsenabled

\commentsenabledtrue

\newcommand\coloruline[1]{\bgroup\markoverwith{\textcolor{#1}{\rule[-0.5ex]{2pt}{1pt}}}\ULon}

\def\BibTeX{{\rm B\kern-.05em{\sc i\kern-.025em b}\kern-.08em
    T\kern-.1667em\lower.7ex\hbox{E}\kern-.125emX}}
\begin{document}

\title{\NAME{}: Using Graphs to Compress Smaller and Faster}

\author{\IEEEauthorblockN{Yann Collet, Nick Terrell, Winston\ Felix Handte, Danielle Rozenblit, Victor Zhang, \\ Kevin Zhang, Yaelle Goldschlag, Jennifer Lee, Elliot Gorokhovsky, Yonatan Komornik, \\ Daniel Riegel, Stan Angelov, Nadav Rotem}
\IEEEauthorblockA{\textit{Meta Platforms, Inc.} \\
\{cyan, terrelln, felixh, drozenblit, csv, kevz8, ygoldschlag, jenlee68\}@meta.com, \\ \{elliot.gorokhovsky, yoniko\}@gmail.com, \{riegel, sangelov, nrotem\}@meta.com
}
}

\maketitle

\begin{abstract}
In the last few decades, research techniques have improved lossless compression ratios by significantly increasing processing time. However, these techniques have not gained popularity in industry because production systems require high throughput and low resource utilization. Instead, real world improvements in compression are increasingly realized by building application-specific compressors which can exploit knowledge about the structure and semantics of the data being compressed. Application-specific compressor systems outperform even the best generic compressors, but these techniques have severe drawbacks---they are inherently limited in applicability, are hard to develop, and are difficult to maintain and deploy.

In this work, we show that these challenges can be overcome with a new compression strategy. We propose the ``graph model'' of compression, a new theoretical framework for representing compression as a directed acyclic graph of modular codecs. OpenZL implements this framework and compresses data into a self-describing wire format, any configuration of which can be decompressed by a universal decoder. OpenZL’s design enables rapid development of application-specific compressors with minimal code; its universal decoder eliminates deployment lag; and its need for a limited set of standard components minimizes maintenance burden. Experimental results demonstrate that OpenZL achieves superior compression ratios and speeds compared to state-of-the-art general-purpose compressors on a variety of real-world datasets. Compared to ratio-focused deep-learning compressors, OpenZL is competitive on ratio while being many orders of magnitude faster. Internal deployments at Meta have also shown consistent improvements in size and/or speed, with development timelines reduced from months to days. OpenZL thus represents a significant advance in practical, scalable, and maintainable data compression for modern data-intensive applications.
\end{abstract}

\begin{IEEEkeywords}
compression, data models, database management, heterogeneous databases, training
\end{IEEEkeywords}

\section{Introduction}
\label{sections:introduction}

Compression research over the last decade has largely focused on leveraging machine learning to improve compression ratios~\cite{KNOLL12, CMIX, NNCP, GOYAL20, LSTM-COMPRESS, MAO22}. This benefits scenarios where minimizing data size is critical but speed is less important~\cite{LIU22, LSTM-COMPRESS, NNCP, ZHANG24}. Deep-learning approaches compress benchmark datasets at significantly better ratios than traditional techniques but achieve throughput on the order of 1KB/s and often require heavy GPU resources~\cite{LIU22, GOYAL20, CMIX, ZHANG24}.

By contrast, production workloads must strike a balance between compressed size and processing time. For this reason, almost all popular compressors on the market use a variant of LZ77~\cite{GUPTA17}, as its fast speeds and reasonable compression ratio makes it suitable for latency-sensitive, high-volume environments. Indeed, the last great leap in production-scale compression was Zstandard (Zstd)~\cite{zstandard}, which combines LZ77 with entropy coding, and whose typical usage compresses on the order of 100 MB/s and decompresses on the order of 1 GB/s. Many papers investigating compression for real-world applications say the quiet part out loud: deep-learning systems are too slow and require too many resources to be serious contenders in production, despite superior ratios~\cite{JUMAR18, THEVENIN22, IQBAL20, 2BrightSparks24, LinuxReviews}.

As a consequence, domain-specific compressors are increasingly the tool of choice for data-intensive workflows. In fields like genomics~\cite{PIBIRI22, PIBIRI23, ALANKO25, CHANDAK18, LAN21, CHANDRA12}, computer graphics~\cite{MENTZER20, TAUBMAN19, Pranckevicius25, MESHOPTIMIZER}, and AI models~\cite{HERSHCOVITCH24}, tailored compression algorithms have pushed the state of the art in both academic and industrial applications.

Across disparate read/write patterns, data lifetime requirements, and data organization, there is a clear through-line: knowing \emph{anything at all} about the data being compressed yields better and faster compression than even the best generic compressors. And furthermore, the more structure that can be exploited out of the data, the better performance will be on both axes.

This then begs the question: why aren't application-specific compressors more common? In other words, if there's a clear way to be both smaller and faster than generic compressors, \emph{why is it so rare?}

\begin{enumerate}[label=(\roman*)]
    \item \textbf{Upfront investment is intractable.} Designing and implementing compression algorithms requires significant expertise and effort. Zstd, for instance, contains over 100,000 lines of code, representing years of development and hand-optimization from a well-funded team. Beyond this, application-specific compression algorithms requires expertise not only in compression techniques, but also in the problem domain.
    \item\label{bad2} \textbf{Inflexibility of solutions.} Most custom compressors are designed to optimize for type of data. Scale issues are immediate once you try to onboard datasets that require different compression techniques. Additionally, just supporting new wire formats of the same data often requires extensive refactoring or even a complete rebuild of the library.
\end{enumerate}

These constraints prevent smaller teams from adopting custom compression solutions. For larger enterprises that can fund their way out of these challenges, a different set of constraints arise when custom solutions are deployed at scale.

\begin{enumerate}[resume*]
    \item \textbf{Iteration is difficult.} Releasing a production library means freezing the wire format and publishing long-term support guidance. Often this inhibits the speed of new development, as backwards compatibility limits what you can ship. This reduces the efficacy of custom compression because not all gains can be realized.
    \item \textbf{Deployment is slow.} Updating from version $n$ to $n + 1$ requires that all the data readers are rolled out to support version $n+1$ before any of the writers are allowed to write the new version. Thus, without guarantees on library freshness, updating the wire format becomes untenable. This makes custom compression unsuitable for whole swaths of applications, including mobile app development and IoT devices.
    \item \textbf{High-cardinality applications are difficult to secure.} For data warehouses with diverse customer needs, deploying custom codecs for each of your clients' needs multiplies the amount of code that needs to be maintained and security-hardened.
\end{enumerate}

In this paper, we show that these challenges can be overcome with a new way of thinking about compression. We introduce \NAME{}, a general compression engine that uses a graph-structured compression model with a self-describing wire format and a universal decoder. Specifically, \NAME{} breaks from the typical monolithic compressor architecture by decomposing a compression into a DAG of composable codecs.

\subsection{The Graph Model of Compression}

Our main theoretical contribution is the graph model of compression, defined formally in \cref{sections:concepts}. In summary, we define a \textit{compression graph} as a computational graph~\cite{COLLINS18} where the nodes are \textit{codecs} and edges represent data generated as output of one codec and used as input for another. \textit{Codecs} are defined simply to be multivariate functions, so are allowed to have multiple inputs and multiple outputs. The graph model allows us to think about compressors as abstract transformation engines and enables new ways of approaching the task of compression.

The computational graph also means decompression is purely procedural. Apart from the compressed data itself, all you need is the graph to decode any valid compressed frame. The universality of the decoder is an important result of the graph model.

\subsection{Overview of Results}

This paper's main result is demonstrating that \NAME{}'s graph model is both easy to use and expressive enough to cover a wide range of applications. \NAME{} is able to address several pain points of application-specific compressors in order to simplify their development:

\begin{enumerate}[label=(\roman*)]
    \item \textbf{Upfront investment is minimal.} Compared to existing monolithic compressor architectures, \NAME{}'s modular structure solves the flexibility problem and allows the same developer to quickly stand up compressors for many different formats. We show in \cref{subsections:compressor-training} that little, if any, production code needs to be written to support new datasets.
    \item \textbf{Solutions are flexible.} The composable graph model also means diverse datasets can be supported by simply creating new graphs. Our benchmark experiments in \cref{sections:results} demonstrate that many disparate data formats can be easily parsed to take advantage of \NAME{}.
\end{enumerate}

\NAME{} is also an enterprise-scale solution for custom compression.

\begin{enumerate}[resume*]
    \item \textbf{Iteration is easy.} The self-describing wire format means you can evolve a compressor's graph over time without needing to make any changes to the decoder.
    \item \textbf{Deployment is fast.} A universal decoder elides the rollout calculation. The same core library can decode any graph given to it.
    \item \textbf{High-cardinality applications are easy to secure.} By centralizing compression onto one library, the maintenance and security surface area does not increase as new use cases are onboarded.
\end{enumerate}

\subsection{Paper Organization}

\Cref{sections:related_work_2} describes prior work on data compression, in particular composable compression engines.

\Cref{sections:concepts} develops the graph model of compression, the underlying theoretical framework for \NAME{}.

\Cref{sections:compressor_generation_sao} provides a worked example, using graph model principles to build a compressor for an example dataset.

\Cref{sections:implementation} explores \NAME{}'s implementation of the graph model and design choices of the code that allows it to take the leap from theory to practice.

\Cref{sections:setup} and \cref{sections:results} describe our experimental setup and results, which show compressors built in \NAME{} that achieve ratio and speed improvements on a variety of different data.

\Cref{sections:openzl_at_meta} details the benefits we've seen from internal deployment at Meta.

\Cref{sections:conclusions} contains some concluding remarks and a look to the future.

\section{Related Work}
\label{sections:related_work_2}

It is well-known that a universal algorithm that always compresses every input does not exist. If some inputs are shortened, others are necessarily lengthened. In other words, lossless compression algorithms are injective, and for injective $f: A \to B$ it holds that $|B| \geq |A|$.

\subsection{Entropy Coding}

It follows that a valid strategy is to fill the space unevenly, assigning short codes to frequent events, while sacrificing rarer ones with longer codes.
This is the core of \textit{entropy coders}, which have been well studied for decades, and offer a predictable ceiling, named the \textbf{Shannon limit}~\cite{SHANNON}.

\textbf{Huffman} remains widely used~\cite{huff}; \textbf{arithmetic coding} approaches the entropy limit with different latency/state trade-offs~\cite{WITTEN87}. The \textbf{ANS} family offers arithmetic-like compression at higher speed~\cite{duda2014}, with Zstd’s FSE~\cite{FSE} as a table-driven tANS example. 

In practice, entropy coders are mature; current work centers on high-throughput, optimized implementations and careful quantization of probabilities to the coder’s precision.

\subsection{Reversible Transforms}

To reach high compression ratios, entropy alone is not enough.
Instead of compressing raw input symbols, a more favorable strategy is to first convert the input into another representation, using some reversible operation. 

Reversible transforms are simply transforms with an inverse operation.
Under this definition, an infinite number of transformations match this criteria.
For example, encryption is technically a reversible operation. But it wouldn't help for data compression.
So it matters to select the correct one for the current context.

A better example would be an increasing numeric series, which can be converted into a series of \textbf{deltas}. Not only will this mechanically reduce the range of represented values, making the signal easier to compress, but from the operation may emerge a regularity (for example, sequential values are transformed into a series of 1s) that will make the resulting signal even easier to compress.

Another example, the \textbf{Burrows--Wheeler Transform (BWT)}~\cite{burrows1994bwt} clusters similar contexts so nearby symbols look alike. The produced stream has the same size and content as the ingested one, but ordered differently. It's often followed by \textbf{move-to-front (MTF) coding}~\cite{MTF}, which reorders symbols by recency index, thus exposing locality for entropy coding.
We can also present \textbf{Transpose}, which reorders bytes according to their rank in a fixed-size multi-bytes field.

Reversible Transformations are beneficial when they help expose structure for downstream modeling. 
Some transforms may produce less data than they ingested. But that's not a rule, and is not necessary for a transform to be useful.

\subsection{Reductive Transforms}

Reductive Transforms are also reversible, but their main contribution is to reduce the amount of data to process by further stages.

For example, \textbf{Run-length encoding (RLE)}~\cite{RLE} collapses symbol runs. The \textbf{LZ77} family~\cite{lz77} replaces repeated substrings with backward references, while the \textbf{LZ78} one~\cite{lz77} replaces them by an index in a dynamic dictionary. Another common strategy is \textbf{dictionary substitution}, where frequent substrings are factored into a static table and the input is rewritten as indices into that table. 

In these cases, substantial reduction in data size is the intended effect, reflected in both final compression ratio and processing speed (having less data to process requires less computing time).

\subsection{General-Purpose LZ}

Faced with so many possibilities, the challenge is to select a proper set of transformation stages.
Most compressors address this challenge by settling for a fixed set of transformation stages deemed "good enough" for general purpose use.

For scenarios with strict throughput requirements, \textbf{LZ4}~\cite{lz4} and \textbf{Snappy}~\cite{snappy} apply LZ-style parsing with a lightweight tagged format (varint lengths/distances) rather than a full entropy-coding stage. They are effective for structured and textual data where modest ratios suffice but (de-)compression speed is critical.

\textbf{DEFLATE/gzip}~\cite{rfc1951, rfc1952} encodes literal/length and distance symbols, intertwined in a single Huffman-coded stream, using (static or dynamic) Huffman trees. \textbf{Brotli}~\cite{rfc7932} improves on gzip by using context models for literals and offsets. \textbf{Zstd}~\cite{rfc8878} factors tokens into four logical streams (literals, literal-lengths, match-lengths, offsets) and uses either Huffman or FSE depending on mode. \textbf{LZMA} (as used in \textbf{xz}) combines LZ parsing and a range coder with context models (it is not a ``pure LZ'' design, making it more powerful but also markedly slower)~\cite{LZMA}.

\subsection{Prediction-Centric Compressors}

Higher compression ratios hinge on accurate symbol prediction. \textbf{PPM} conditions on preceding contexts~\cite{ppm}; \textbf{DMC} learns an adaptive automaton~\cite{dmc}; \textbf{context mixing} (e.g., \textbf{PAQ}~\cite{mahoney2013paq}, \textbf{cmix}~\cite{CMIX}) blends multiple predictors using ideas related to boosting. Today these are increasingly neural-net based. \textbf{NNCP} is a more recent approach in this vein~\cite{NNCP}. Despite excellent ratios, these algorithms remain orders of magnitude slower (KB/s scale) and sequential, making them less attractive for datacenter hot paths than the much faster LZ options.

\subsection{A Programmable Composition of Processing Stages}

Staged pipeline designs are explicit in several widely used formats. \textbf{PNG}~\cite{rfc2083} applies a per-scanline prediction filter (None/Sub/Up/Average/Paeth) before DEFLATE, turning image structure into locally predictable residuals that the downstream coder compresses well; the filter choice is itself a stage decision recorded per row. 
\textbf{Blosc}~\cite{BLOSC} composes blocking, a shuffle/bitshuffle transform, and then a configurable backend codec (LZ4, Zstd, etc.) 
\textbf{Parquet}~\cite{PARQUET} composes per-column encodings (dictionary, delta, RLE/bit-packing) with a backend codec. 
In both of these cases, the composition is tunable. Indeed, Blosc offers \textbf{BTune}~\cite{BTUNE}, which explores automatic choice/parameter search across Blosc2 codecs and filters. 
\textbf{ZPAQ} goes further: the archive stores a virtual-machine program~\cite{mahoney2015zpaq} specifying contexts/transforms and the coder. Here, the pipeline is part of the compressed frame.

In practice, these designs are constrained by enumerated stage catalogs and fixed rules on how these stages can be composed. Even where plugins exist, tuning often focuses on parameters for individual stages rather than exploring different stage orderings or richer graphs. As a result, a large fraction of the transform design space---and the cross-stage interactions that dictate effectiveness---remains unexplored.

\NAME{} represents the natural evolution of the staged pipeline design. Rather than limit the system to a strict linear flow, we introduce the graph model of compression. 

\section{The Graph Model of Compression}
\label{sections:concepts}

In this section we develop the abstract graph model. In the next section, we build a sample compressor using the graph model.

\subsection{Data and Messages}

In typical data compression parlance, a \textit{message} is a sequence of bytes. In the graph model, we adopt a stricter requirement for messages.

\begin{definition}[Message Sets]
    A \textbf{message set} is a non-empty subset of the universe of bitstrings.
\end{definition}

Under this framing, a message is an element drawn from a message set. Rather than be any random bitstring, we impose semantic requirements on messages by restricting the possibility set. For instance, components may require that messages represent 64-bit integer arrays by requiring that all messages have bit-length divisible by 64. More restrictive representational requirements can also be imposed. For instance, sorted runs of bytes; or a particular semantic structure like a zip file.

\subsection{Codecs}

Fundamentally, a codec is little more than a function operating on message sets.

\begin{definition}[Codec]
An input (resp. output) is an ordered tuple of messages $\mu = (\mu_1, \dots, \mu_n)$, each drawn from a potentially different message set $\mu_i \in X_i$. The \textbf{input domain} (resp. \textbf{output domain}) is the ordered tuple of these message sets, i.e. $X = (X_1, \dots, X_n)$.
A \textbf{codec} is a tuple $(C,D)$ of functions. The \textit{encoder} $C: I \to O$ is a mapping between a non-empty input domain $I$ and a non-empty output domain $O$. The \textit{decoder} $D: O \to I'$ maps $O$ to a possibly different regenerated domain $I'$. A codec is \textit{lossless} if this mapping is invertible, that is, $I \equiv I'$ and
$$D(C(\mu)) \equiv \mu, \; \forall \mu \in I \;.$$
\end{definition}

This definition intentionally de-emphasizes the inner workings of the codec. In this model, the input/output signature matters more than the exact implementation because the signature tells us how the information is transformed semantically. This is an important abstraction, as it enables us to consider composition at a higher level.

\begin{remark}
    From an implementation perspective, codecs are most useful when they are small and limited in scope. \Cref{sections:implementation} describes the implementation philosophy in \NAME{}.
\end{remark}

\subsection{Composition and Graphs}

In the graph model, compressors are graphs built from codec nodes. As a motivating example, consider the \texttt{tokenize} codec. Briefly, \texttt{tokenize} searches for repeated instances of the same ``token''. It works by taking a message $\mu \in \Sigma^*$ and outputting 2 messages: $\alpha$, the list of unique tokens in $\mu$; and $\nu$, the ``index'' in $\alpha$ of each token within $\mu$.

\begin{figure}[H]
    \centering
    \footnotesize
    \begin{tikzpicture}[]
    \node [shape=rectangle,fill=orange!20, align=center](table1) at (0.0,0.0) {
            $\mu$ \\
            \begin{tabular}{c|c|c|c|c|c|c} \toprule
                \verb|alice| & \verb|bob| & \verb|bob| & \verb|eve| & \verb|alice| & \verb|bob| & \verb|alice|\\
                \bottomrule
            \end{tabular}
        };
    \node [shape=rectangle,fill=blue!20, align=center](table2) at (-1.0,-2.0) {
            $\nu$ \\
            \begin{tabular}{c|c|c|c|c|c|c} \toprule
                0 & 1 & 1 & 2 & 0 & 1 & 0\\
                \bottomrule
            \end{tabular}
        };
    \node [shape=rectangle,fill=blue!20, align=center](table3) at (3.0,-2.0) {
            $\alpha$ \\
            \begin{tabular}{c} \toprule
                \verb|alice|\\ \midrule
                \verb|bob|\\ \midrule
                \verb|eve|\\
                \bottomrule
            \end{tabular}
        };
    \draw [thin, -Stealth] (table1) -- (table2);
    \draw [thin, -Stealth] (table1) -- (table3);
\end{tikzpicture}
    \caption{An example invocation of the \texttt{tokenize} codec.}
    \label{fig:tokenize}
\end{figure}
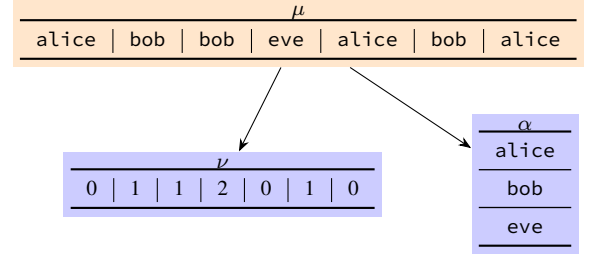

Tokenization is sometimes an effective compressor on its own (such as when the message is composed of many repetitions of a few large tokens). More frequently though, its utility is as an intermediate transformation, which produces outputs better suited for subsequent processing. We can attach other codecs to each of the \texttt{tokenize} codec's two outputs, which separately attack the problems of efficiently representing the contents of the tokens and the indices.

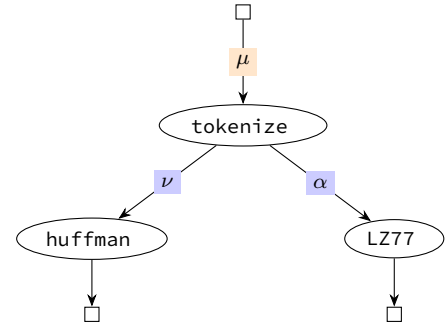
\begin{figure}[H]
    \centering
    \footnotesize
    \begin{tikzpicture}[]
    \node [shape=rectangle, draw, fill=white, align=center](ROOT) at (2.0,1.5) {};
    \node [shape=ellipse, draw, fill=white, align=center](TOK) at (2.0,0.0) {
        \verb|tokenize|
    };
    \node [shape=ellipse, draw, fill=white, align=center](HUFF) at (0.0,-1.5) {
            \verb|huffman|
    };
    \node [shape=ellipse, draw, fill=white, align=center](LZ77) at (4.0,-1.5) {
        \verb|LZ77|
    };
    \node [shape=rectangle, draw, fill=white, align=center](store1) at (0.0, -2.5) {};
    \node [shape=rectangle, draw, fill=white, align=center](store2) at (4.0, -2.5) {};
    \draw [thin, -Stealth] (ROOT) -- (TOK) node [midway, fill=orange!20] {$\mu$};
    \draw [thin, -Stealth] (TOK) -- (HUFF) node [midway, fill=blue!20] {$\nu$};
    \draw [thin, -Stealth] (TOK) -- (LZ77) node [midway, fill=blue!20] {$\alpha$};
    \draw [thin, -Stealth] (HUFF) -- (store1);
    \draw [thin, -Stealth] (LZ77) -- (store2);
\end{tikzpicture}
    \caption{An example compressor that uses tokenize, Huffman, and LZ77.}
    \label{fig:tokenize_compressor}
\end{figure}

While the alphabet $\alpha$ has the same type of content as the original message $\mu$, the indices in $\nu$ are a sequence of integers rather than strings. While $\nu$ is a partial representation of $\mu$, by transforming it into a fixed-width, integer sequence, we can bring techniques to bear on it that can't be applied directly to $\mu$. For instance, we may construct a compressor that sends $\alpha$ to an LZ77 compressor and $\nu$ to an entropy encoder like Huffman. \Cref{fig:tokenize_compressor} contains a visualization of this new compressor.

As the visualization implies, a compressor with multiple codecs conveniently organizes itself as a graph. Informally, a \textbf{compression graph} is a computational graph~\cite{COLLINS18, DYER16} where the nodes represent codecs and edges represent input and output sets\footnote{Technically, this is a \emph{reversed} computational graph, since in typical depictions the feed-forward direction merges multiple inputs to produce the output, whereas a compression graph generates multiple outputs from the input.} In particular, an edge between a parent and child codec indicates a sequential relationship, where one of the outputs of the parent is used as one of the inputs of the child.

\begin{definition}[Compression Graph]
    Formally, a \textbf{computational graph} is a directed, acyclic, graph (DAG) where the nodes are functions and the edges represent function arguments (and data dependencies).
    A \textbf{compression graph} is a computational graph where each node $v$ is labelled with a codec $C_v: I_v \to O_v$, and edges $u \to v$ are doubly-labelled with both an output from the source $(O_u)_i$ and an input to the target $(I_v)_j$ such that $(O_u)_i \subseteq (I_v)_j$.
\end{definition}

The sequence of transforms permitted by this model allows us to build compressors that exploit the semantics of the data much better than generic compressors. Semantic specialization in intermediate streams increases as you traverse the compression graph, increasing the efficiency of subsequent codecs. For entropy coders, such specialization can result in intermediate representations with lower entropy and thus a more compact code.

\subsection{Universal Decoder}

The compression graph inherits some useful properties from computational graphs. Notably, computational graphs are DAGs. And since every DAG admits a topological sort, this ensures that a well-defined compression graph always admits a valid feed-forward computation order (for compression) and a valid backpropagation order (for decompression).

Decompression thus operates by procedurally chaining together codec decoders in the order dictated by the topological sort.
Moreover, since the proper decode procedure is uniquely determined by a combination of the final outputs and the graph structure, a graph-based compressor can exploit this property to provide a universal decoder, assuming it can decode all the codecs being used.

\subsection{Compression Dynamicity}

As presented, the graph model does not allow for any dynamism in the choice of codecs to run in a graph. However, strong guarantees on decodability motivates us to add flexibility with a small extension to the model.

\begin{definition}[Resolved Graph]
    A \textbf{resolved graph} is a compression graph that contains only codecs.
\end{definition}

The example graph in \cref{fig:tokenize_compressor} is a resolved graph, as are all the graphs we have discussed so far.

\begin{definition}[Function graph]
    Let $\mathbb{G}$ denote the set of compression graphs. A \textbf{function graph} is a function $F: I \to \mathbb{G}$. A \textbf{dynamic compression graph} is a compression graph where the nodes are either codecs or function graphs.
\end{definition}

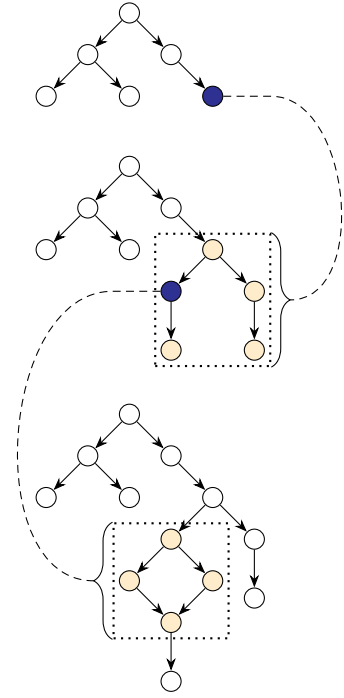
\begin{figure}[h]
    \centering
    \footnotesize
    \begin{tikzpicture}[
  node/.style={circle, draw},
  edge/.style={-Stealth},
]

\node[node,                      ] (A1) {};
\node[node, below left =0.5 of A1] (B1) {};
\node[node, below right=0.5 of A1] (C1) {};
\node[node, below left =0.5 of B1] (D1) {};
\node[node, below right=0.5 of B1] (E1) {};
\node[node, below right=0.5 of C1, fill=Blue] (F1) {};

\draw[edge] (A1) -- (B1);
\draw[edge] (A1) -- (C1);
\draw[edge] (B1) -- (D1);
\draw[edge] (B1) -- (E1);
\draw[edge] (C1) -- (F1);

\node[node, below      =1.75 of A1] (A2) {};
\node[node, below left =0.5  of A2] (B2) {};
\node[node, below right=0.5  of A2] (C2) {};
\node[node, below left =0.5  of B2] (D2) {};
\node[node, below right=0.5  of B2] (E2) {};
\node[node, below right=0.5  of C2, fill=Dandelion!25] (F2) {};
\node[node, below left =0.5  of F2, fill=Blue] (G2) {};
\node[node, below right=0.5  of F2, fill=Dandelion!25] (H2) {};
\node[node, below      =0.5  of G2, fill=Dandelion!25] (I2) {};
\node[node, below      =0.5  of H2, fill=Dandelion!25] (J2) {};
\draw[edge] (A2) -- (B2);
\draw[edge] (A2) -- (C2);
\draw[edge] (B2) -- (D2);
\draw[edge] (B2) -- (E2);
\draw[edge] (C2) -- (F2);
\draw[edge] (F2) -- (G2);
\draw[edge] (F2) -- (H2);
\draw[edge] (G2) -- (I2);
\draw[edge] (H2) -- (J2);

\node [draw, thick, dotted, fit=(F2) (G2) (H2) (I2) (J2), inner sep=2pt] (EXPANSION2) {};

\draw [decorate,decoration={brace,amplitude=0.25cm}]
      (EXPANSION2.north east)
      -- node [midway] (EXPANSION2TOP) {}
      (EXPANSION2.south east);

\draw [densely dashed] (F1) to (EXPANSION2TOP|-F1) to [out=  0, in=  0] ($(EXPANSION2TOP)+(0.25,0)$);

\node[node, below      =3   of A2] (A3) {};
\node[node, below left =0.5 of A3] (B3) {};
\node[node, below right=0.5 of A3] (C3) {};
\node[node, below left =0.5 of B3] (D3) {};
\node[node, below right=0.5 of B3] (E3) {};
\node[node, below right=0.5 of C3] (F3) {};
\node[node, below left =0.5 of F3, fill=Dandelion!25] (G3) {};
\node[node, below right=0.5 of F3] (H3) {};
\node[node, below left =0.5 of G3, fill=Dandelion!25] (I3) {};
\node[node, below right=0.5 of G3, fill=Dandelion!25] (J3) {};
\node[node, below right=0.5 of I3, fill=Dandelion!25] (K3) {};
\node[node, below      =0.5 of K3] (L3) {};
\node[node, below      =0.5 of H3] (M3) {};
\draw[edge] (A3) -- (B3);
\draw[edge] (A3) -- (C3);
\draw[edge] (B3) -- (D3);
\draw[edge] (B3) -- (E3);
\draw[edge] (C3) -- (F3);
\draw[edge] (F3) -- (G3);
\draw[edge] (F3) -- (H3);
\draw[edge] (G3) -- (I3);
\draw[edge] (G3) -- (J3);
\draw[edge] (I3) -- (K3);
\draw[edge] (J3) -- (K3);
\draw[edge] (K3) -- (L3);
\draw[edge] (H3) -- (M3);

\node [draw, thick, dotted, fit=(G3) (I3) (J3) (K3), inner sep=2pt] (EXPANSION3) {};

\draw [decorate,decoration={brace,amplitude=0.25cm}]
      (EXPANSION3.south west)
      -- node [midway] (EXPANSION3TOP) {}
      (EXPANSION3.north west);

\draw [densely dashed] (G2) to (EXPANSION3TOP|-G2) to [out=180, in=180] ($(EXPANSION3TOP)+(-0.25,0)$);

\end{tikzpicture}
    \caption{An example of function graph expansion. Function graphs are shaded and their expansions marked in dotted lines. }
    \label{fig:function_graph}
\end{figure}

Function graphs can be described as ``selectors'', choosing a graph based on the input, which itself may contain additional function graphs. At compression time, this expansion naturally modifies the graph being run. \Cref{fig:function_graph} illustrates this process.

Readers familiar with lambda calculus may draw a vague parallel between function graph expansion and beta-reduction. Like beta-reduction, the function graph expansion process creates another valid graph, which may have more opportunities for function graph expansion. The ``beta-normal form'' for function graphs is the resolved graph, wherein no more expansion is possible.

A compression that succeeds will always generate a resolved graph. Since the resolved graph contains only regular codecs, it also completely specifies how to reconstruct the original input. In the graph model, the decoder cannot make any runtime decisions based on data presented, so the guaranteed existence of a resolved graph allows us to safely incorporate dynamism into \NAME{}.

\section{Compressor generation by example}
\label{sections:compressor_generation_sao}

The freedom of composition allowed by the graph model can be daunting. Faced with limitless ways to interpret data, it matters to introduce methodologies to structure the effort. We present the following method:

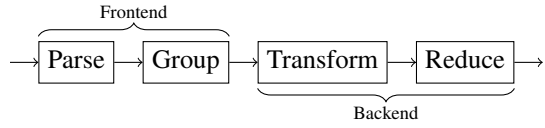
\begin{figure}[H]
    \centering
    \begin{tikzpicture}
\node [draw, fill=white] (PARSE) {Parse\vphantom{jk}};
\node [draw, fill=white, right=0.375 of PARSE] (GROUP) {Group\vphantom{jk}};
\node [draw, fill=white, right=0.375 of GROUP] (TRANSFORM) {Transform\vphantom{jk}};
\node [draw, fill=white, right=0.375 of TRANSFORM] (REDUCE) {Reduce\vphantom{jk}};

\draw [->] ($(PARSE.west)+(-0.375,0)$) -- (PARSE);
\draw [->] (PARSE) -- (GROUP);
\draw [->] (GROUP) -- (TRANSFORM);
\draw [->] (TRANSFORM) -- (REDUCE);
\draw [->] (REDUCE) -- ($(REDUCE.east)+(0.375,0)$);

\draw [decorate, decoration={brace,amplitude=5pt,raise=0.0625cm}]
      (PARSE.north west)
      -- node [midway, above=5pt] {\scriptsize Frontend}
      (GROUP.north east);

\draw [decorate, decoration={brace,amplitude=5pt,raise=0.0625cm}]
      (REDUCE.south east)
      -- node [midway, below=5pt] {\scriptsize Backend}
      (TRANSFORM.south west);
\end{tikzpicture}
    \label{fig:compressor_structure}
    \caption{Common abstract compressor structure.}
\end{figure}


This pattern emerges because the components of \NAME{} that are actually good at compressing data---its suite of reversible/reductive transforms---work best on homogeneous streams of data. For inputs that aren't already organized that way, those backend components require a frontend to parse and group the input into streams that the backend can then compress effectively.

To illustrate this pattern, we will use SAO, a part of the Silesia Compression Corpus~\cite{SILESIA}. This file follows a well-documented format~\cite{SAO} featuring a small header followed by an array of multi-fields records, each one describing a star.

\subsubsection{Frontend}
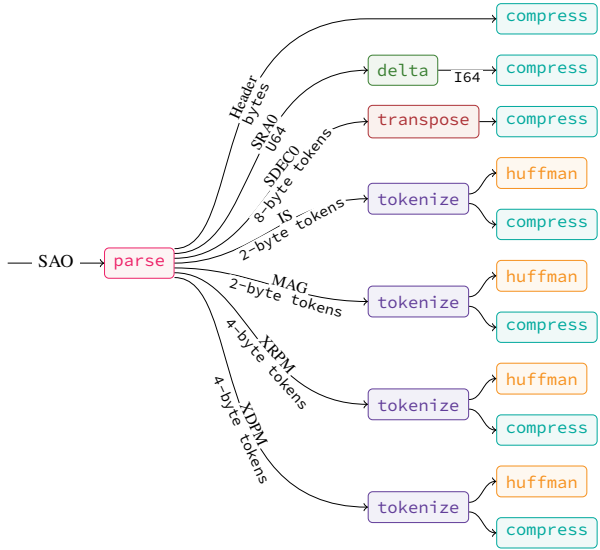
\begin{figure}[h]
    \centering
    \footnotesize
    \scalebox{0.85}{\begin{tikzpicture}[global]
\node [sao transform, Emerald, fill=Emerald!5!white] (HEADER_COMPRESS) {\tt compress};
\node [sao transform, below=.8 of HEADER_COMPRESS.west, anchor=west, Emerald, fill=Emerald!5!white] (SRA0_COMPRESS) {\tt compress};
\node [sao transform, below=.8 of SRA0_COMPRESS.west, anchor=west, Emerald, fill=Emerald!5!white] (SDEC0_COMPRESS) {\tt compress};
\node [sao transform, below=.8 of SDEC0_COMPRESS.west, anchor=west, BurntOrange, fill=BurntOrange!5!white] (IS_HUFF) {\tt huffman};
\node [sao transform, below=.8 of IS_HUFF.west, anchor=west, Emerald, fill=Emerald!5!white] (IS_COMPRESS) {\tt compress};
\coordinate (IS_CENTER) at ($(IS_HUFF.west)!.5!(IS_COMPRESS.west)$);
\node [sao transform, below=.8 of IS_COMPRESS.west, anchor=west, BurntOrange, fill=BurntOrange!5!white] (MAG_HUFF) {\tt huffman};
\node [sao transform, below=.8 of MAG_HUFF.west, anchor=west, Emerald, fill=Emerald!5!white] (MAG_COMPRESS) {\tt compress};
\coordinate (MAG_CENTER) at ($(MAG_HUFF.west)!.5!(MAG_COMPRESS.west)$);
\node [sao transform, below=.8 of MAG_COMPRESS.west, anchor=west, BurntOrange, fill=BurntOrange!5!white] (XRPM_HUFF) {\tt huffman};
\node [sao transform, below=.8 of XRPM_HUFF.west, anchor=west, Emerald, fill=Emerald!5!white] (XRPM_COMPRESS) {\tt compress};
\coordinate (XRPM_CENTER) at ($(XRPM_HUFF.west)!.5!(XRPM_COMPRESS.west)$);
\node [sao transform, below=.8 of XRPM_COMPRESS.west, anchor=west, BurntOrange, fill=BurntOrange!5!white] (XDPM_HUFF) {\tt huffman};
\node [sao transform, below=.8 of XDPM_HUFF.west, anchor=west, Emerald, fill=Emerald!5!white] (XDPM_COMPRESS) {\tt compress};
\coordinate (XDPM_CENTER) at ($(XDPM_HUFF.west)!.5!(XDPM_COMPRESS.west)$);
\node [sao transform, left=2 of SRA0_COMPRESS.west, anchor=west, OliveGreen, fill=OliveGreen!5!white] (SRA0_DELTA) {\tt delta};
\node [sao transform, left=2 of SDEC0_COMPRESS.west, anchor=west, Maroon, fill=Maroon!5!white] (SDEC0_TRANSPOSE) {\tt transpose};
\node [sao transform, left=2 of IS_CENTER, anchor=west, RoyalPurple, fill=RoyalPurple!5!white] (IS_TOK) {\tt tokenize};
\node [sao transform, left=2 of MAG_CENTER, anchor=west, RoyalPurple, fill=RoyalPurple!5!white] (MAG_TOK) {\tt tokenize};
\node [sao transform, left=2 of XRPM_CENTER, anchor=west, RoyalPurple, fill=RoyalPurple!5!white] (XRPM_TOK) {\tt tokenize};
\node [sao transform, left=2 of XDPM_CENTER, anchor=west, RoyalPurple, fill=RoyalPurple!5!white] (XDPM_TOK) {\tt tokenize};
\node [fit=(HEADER_COMPRESS) (SRA0_COMPRESS) (SDEC0_COMPRESS) (IS_HUFF) (IS_COMPRESS) (MAG_HUFF) (MAG_COMPRESS) (XRPM_HUFF) (XRPM_COMPRESS) (XDPM_HUFF) (XDPM_COMPRESS) (SRA0_DELTA) (SDEC0_TRANSPOSE) (IS_TOK) (MAG_TOK) (XRPM_TOK) (XDPM_TOK), inner sep=0.25cm] (ALL_TRANSFORMS) {};
\node [fit=(HEADER_COMPRESS) (SRA0_DELTA) (SDEC0_TRANSPOSE) (IS_TOK) (MAG_TOK) (XRPM_TOK) (XDPM_TOK), inner sep=0pt] (FIRST_TRANSFORMS) {};
\node [sao transform, left=3 of FIRST_TRANSFORMS.west, WildStrawberry, fill=WildStrawberry!5!white] (PARSER) {\tt parse};
\node [left=1.5 of PARSER] (SAO) {};
\draw [->]
      (SAO)
      --
      node [midway, fill=white, txt, fn] {SAO}
      (PARSER);
\draw [->]
      ([yshift= .225cm]PARSER.east)
      to [out=  0, in=180, out looseness=0.6, in looseness=1.4]
      node [pos=0.6, sloped, sao stream node, above=-1pt] {Header}
      node [pos=0.6, sloped, sao stream node, below=0pt, tiny] {\tt bytes}
      (FIRST_TRANSFORMS.west|-HEADER_COMPRESS)
      to
      (HEADER_COMPRESS);
\draw [->]
      ([yshift= .150cm]PARSER.east)
      to [out=  0, in=180, out looseness=0.8, in looseness=1.2]
      node [pos=0.6, sloped, sao stream node, above=-1pt] {SRA0}
      node [pos=0.6, sloped, sao stream node, below=0pt, tiny] {\tt U64}
      (FIRST_TRANSFORMS.west|-SRA0_DELTA)
      to
      (SRA0_DELTA);
\draw [->]
      ([yshift= .075cm]PARSER.east)
      to [out=  0, in=180, out looseness=1.0, in looseness=1.0]
      node [pos=0.6, sloped, sao stream node, above=-1pt] {SDEC0}
      node [pos=0.6, sloped, sao stream node, below=0pt, tiny] {\tt 8-byte tokens}
      (FIRST_TRANSFORMS.west|-SDEC0_TRANSPOSE)
      to
      (SDEC0_TRANSPOSE);
\draw [->]
      ([yshift= .0  cm]PARSER.east)
      to [out=  0, in=180]
      node [pos=0.6, sloped, sao stream node, above=-1pt] {IS}
      node [pos=0.6, sloped, sao stream node, below=0pt, tiny] {\tt 2-byte tokens}
      (FIRST_TRANSFORMS.west|-IS_TOK)
      to
      (IS_TOK);
\draw [->]
      ([yshift=-.075cm]PARSER.east)
      to [out=  0, in=180, out looseness=1.0, in looseness=1.0]
      node [pos=0.6, sloped, sao stream node, above=-1pt] {MAG}
      node [pos=0.6, sloped, sao stream node, below=0pt, tiny] {\tt 2-byte tokens}
      (FIRST_TRANSFORMS.west|-MAG_TOK)
      to
      (MAG_TOK);
\draw [->]
      ([yshift=-.150cm]PARSER.east)
      to [out=  0, in=180, out looseness=0.8, in looseness=1.2]
      node [pos=0.6, sloped, sao stream node, above=-1pt] {XRPM}
      node [pos=0.6, sloped, sao stream node, below=0pt, tiny] {\tt 4-byte tokens}
      (FIRST_TRANSFORMS.west|-XRPM_TOK)
      to
      (XRPM_TOK);
\draw [->]
      ([yshift=-.225cm]PARSER.east)
      to [out=  0, in=180, out looseness=0.6, in looseness=1.4]
      node [pos=0.6, sloped, sao stream node, above=-1pt] {XDPM}
      node [pos=0.6, sloped, sao stream node, below=0pt, tiny] {\tt 4-byte tokens}
      (FIRST_TRANSFORMS.west|-XDPM_TOK)
      to
      (XDPM_TOK);
\draw [->]
      (SRA0_DELTA)
      --
      node [sao stream node, midway, below=0pt, tiny] {\tt I64}
      (SRA0_COMPRESS);
\draw [->] (SDEC0_TRANSPOSE) -- (SDEC0_COMPRESS);
\draw [->]
      ([yshift= .075cm]IS_TOK.east)
      to [out=  0, in=180]
      (IS_HUFF);
\draw [->]
      ([yshift=-.075cm]IS_TOK.east)
      to [out=  0, in=180]
      (IS_COMPRESS);
\draw [->]
      ([yshift= .075cm]MAG_TOK.east)
      to [out=  0, in=180]
      (MAG_HUFF);
\draw [->]
      ([yshift=-.075cm]MAG_TOK.east)
      to [out=  0, in=180]
      (MAG_COMPRESS);
\draw [->]
      ([yshift= .075cm]XRPM_TOK.east)
      to [out=  0, in=180]
      (XRPM_HUFF);
\draw [->]
      ([yshift=-.075cm]XRPM_TOK.east)
      to [out=  0, in=180]
      (XRPM_COMPRESS);
\draw [->]
      ([yshift= .075cm]XDPM_TOK.east)
      to [out=  0, in=180]
      (XDPM_HUFF);
\draw [->]
      ([yshift=-.075cm]XDPM_TOK.east)
      to [out=  0, in=180]
      (XDPM_COMPRESS);
\end{tikzpicture}}
    \caption{The simple graph for SAO.}
    \label{figure:sao_graph}
\end{figure}

The \textit{parser} takes the input stream and separates the data into its logical components. For SAO, each record contains 6 fields. The parser turns the array of records into 6 arrays, each containing the concatenated data from the respective field (pictured in \cref{figure:sao_graph}). Including the header, which is passed as-is, the parser produces 7 total streams.

After parsing, \textit{grouping} of fields is sometimes necessary to take full advantage of cross-field correlation. This involves partitioning the parsed streams and merging each partition into a single stream for downstream processing. The simplest (but by no means only) way to do this is to concatenate all streams in a partition. This is overkill for SAO but we use this technique extensively in the experiments (\cref{subsections:compressor-training}).

\subsubsection{Backend}

Now that each stream contains homogeneous data, we can focus on selecting a dedicated compression strategy for each stream. This means selecting a series of reversible/reductive codecs to maximize compression within a given speed budget. This stage is where compression expertise is particularly valuable.

In the case of SAO, we can manually take decisions based on visible characteristics:

\begin{itemize}
    \item \texttt{SRA0} is a position on the X axis. Due to the way the table is generated, the index is mostly sorted, inviting the use of \texttt{delta} to reduce the range of values represented. This mechanically reduces the entropy of the resulting stream, making it easier to compress.
    \item \texttt{SDEC0} is a position on the Y axis. It’s not sorted, unlike the X axis, but we can at least exploit the fact that it’s bounded between a minimum and a maximum. This makes the higher bytes predictable. This can be exploited for better compression with the \texttt{transpose} operation.
    \item The other fields (\texttt{IS}, \texttt{MAG}, \texttt{XRPM}, \texttt{XDPM}) share a common property: their cardinality is much smaller than their quantities, and there is no correlation between consecutive values. This makes them a good target for \texttt{tokenize}, described earlier.
    \item The resulting dictionaries and index lists have very different characteristics. Both are numeric, but one is sparse, the other is dense and bounded. Therefore, they benefit from different compression strategies. They are pushed into different processing graphs.
\end{itemize}


As shown in \cref{table:sao}, this simple analysis and manual decision is enough to produce a compressor\footnote{The full implementation is at \url{https://github.com/facebook/openzl/blob/icde26/cli/utils/compress_profiles.cpp\#L25-L98}} with both a higher compression ratio and faster speed than generic lossless compressors.

\begin{table}[ht]
    \caption{Compression of SAO on an M1 Mac + clang-17}
    \label{table:sao}
    \centering
    \begin{NiceTabular}{l c c c}
    \CodeBefore
    \rectanglecolor{metabg}{1-4}{5-4}
    \Body
    \toprule
     & \texttt{zstd -3} & \texttt{xz -9} & \texttt{OpenZL}\\
    \midrule
    Compressed size (MiB) & 5.28 & 4.21 & 3.35\\
    Compression ratio & 1.31 & 1.64 & 2.06\\
    Compression speed (MiB/s) & 210 & 3.34 & 324\\
    Decompression speed (MiB/s) & 811 & 42.9 & 1140\\
    \bottomrule  
    \end{NiceTabular}
\end{table}

We present this manual example to intuitively illustrate how compression works and how graph choices materially impact compression performance. This could be pushed further---there are more complex relations in the data streams---but exploiting them increases graph complexity. Even with the framework provided by \NAME{}, manually optimizing compressors requires substantial expertise. However, we demonstrate in \cref{subsections:compressor-training} that this process can also be automated to yield good results, making OpenZL performance accessible to non-experts.

\section{Implementation}
\label{sections:implementation}

We now turn from theory to briefly discuss \NAME{}'s implementation of the graph model. This section provides an overview of the structure and design principles of the current open-source implementation, available at \url{github.com/facebook/openzl}.

\subsection{Implementing the Graph Model}

\NAME{} first and foremost is an implementation of the graph model presented in \cref{sections:concepts}. Much of the implementation is unremarkable from a theoretical perspective. We mention some notable exceptions here.
\begin{description}
    \item[Message Sets.] \NAME{} has a partial implementation of message sets. It would be unrealistic to allow specifying arbitrarily-specific sets, so we approximate it with a type system. There are currently 4 types:
    \begin{itemize}
        \item \texttt{bytes} for opaque serial data.
        \item \texttt{string} for sequences of byte strings.
        \item \texttt{struct(k)} for fixed-size (\(k \ge 1\)) records.
        \item \texttt{numeric(w)} a specialization of \texttt{struct} for host-endian 8, 16, 32, and 64-bit numbers.
    \end{itemize}
    \item[Codecs.] Codecs are the lowest level in the OpenZL architecture. Each codec does one thing well. Codecs are split into two parts: encoder and decoder. Implementation-wise, each side is typically organized into two layers: a \emph{kernel} and a \emph{binding}. Kernels are small, deterministic, and allocation free; the binding around them handles types, bounds, and buffers. The vast majority of CPU time is spent in the kernel, so splitting in this way simplifies performance optimization work.
    \item[Dynamism.] We implement dynamism two ways. The \textit{function graph} implements its conceptual namesake: these are regular codec encoders with the restriction that they cannot modify the input data, only call other codecs. The \textit{selector}'s job is to output a graph based on input data. This provides a useful compromise between implementation power and abstract correctness.
\end{description}

\subsection{The Software Stack}

Novelties aside, \NAME{} is fundamentally still an enterprise-scale compression library, so care was taken to design a product conducive to widespread deployment. The open-source OpenZL implementation is layered, with the goal of making the hot path efficient and verifiable.
At the bottom sits the C11 core library, \texttt{libopenzl}, which exposes a stable surface and an execution engine for compression graphs.
A thin C++ façade wraps that surface to provide RAII and strong typing. On top, a Python binding offers an API that integrates with data-science toolchains while preserving the core library's determinism and memory discipline.

The choice of C11 for the core is deliberate: The wide portability and ubiquity of C toolchains, predictable memory semantics, universal ABI, and clear debuggability expectations  are critical at this layer. By constraining the core to a narrow contract, higher layers can evolve at their own cadence without perturbing the on-wire format or the decoder.

\subsection{Versioning and Decoding}

In addition to software ergonomics, \NAME{} addresses the problem of artifact compatibility within its design. For any continuously-deployed software ecosystem, it is inevitable that there are multiple versions of decoders are active at the same time. The traditional solution is to freeze the wire format so all decoders work off the same spec but this limits the evolution of the library. 

\NAME{} uses the concept of \textit{format versions} to add flexibility. When a library version is released, it explicitly supports a range of format versions. At compression time, you select a format version that all your decoders support. Based on the selected version, the library will restrict the suite of functionality it can deploy during the compression. For instance, it can refuse to use new codecs that older formats do not support. In this way, the graph model naturally facilitates incremental binary evolution by breaking down wire-format evolution into a codec-by-codec process.


\subsection{Iteration and Deployment}
All \NAME{} compressors are serializable using any number of graph representation schemes. These \textit{serialized compressors} are very compact; for instance, the SAO  example in the previous section serializes to <2KB. \NAME{} is able to parse (and then compress) with these serialized compressors, so serialized compressors can be passed around and deployed like regular config files.

\section{Experimental Setup}
\label{sections:setup}

In the next two sections, we demonstrate the flexibility and efficacy of \NAME{} by building competitive compressors for a variety of datasets and data formats.

\subsection{Datasets} \label{subsec:standard-compression}

We evaluated the compression ratio of \NAME{} on a number of publicly available datasets. Procedurally, we chose datasets with a variety of file formats. \Cref{table:datasets-summary} summarizes these datasets.

\begin{table}
    \caption{A Summary of Benchmark Datasets Used}
    \label{table:datasets-summary}
    \centering
    \begin{NiceTabular}{ l p{5cm} c c c}
    \CodeBefore
    \Body
    \toprule
    \multicolumn{1}{l}{Dataset} & \multicolumn{2}{r}{Data Format} & Chunked & Mean File Size (MiB)\\
    \midrule
    \multicolumn{2}{l}{\texttt{binance}} & Parquet & No & 6.40\\
    \multicolumn{2}{l}{\texttt{tlc}} & Parquet & No & 248\\
    \multicolumn{2}{l}{\texttt{era5\_flux}} & GRIB & Yes & 7.92\\
    \multicolumn{2}{l}{\texttt{era5\_precip}} & GRIB & Yes & 7.92\\
    \multicolumn{2}{l}{\texttt{era5\_pressure}} & GRIB & Yes & 7.92\\
    \multicolumn{2}{l}{\texttt{era5\_snow}} & GRIB & Yes & 7.92\\
    \multicolumn{2}{l}{\texttt{era5\_wind}} & GRIB & Yes & 7.92\\
    \multicolumn{2}{l}{\texttt{ppmf\_person}} & CSV & Yes & 100\\
    \multicolumn{2}{l}{\texttt{ppmf\_unit}} & CSV & Yes & 100\\
    \multicolumn{2}{l}{\texttt{psam\_h}} & CSV & No & 77.2\\
    \multicolumn{2}{l}{\texttt{psam\_p}} & CSV & No & 173\\
    \bottomrule
    \end{NiceTabular}
\end{table}

\subsubsection{Binance}
An important resource in quantitative finance is candlestick data, which describe how the price of an asset changes over a given timeframe. The Binance dataset~\cite{BINANCE} is a collection of 1-minute candlestick data for the top 1000 cryptocurrency trading pairs on \url{binance.com}, as retrieved from Binance’s official API endpoint for historical candlestick data.

For our benchmark, we selected 15 Bitcoin candlestick records from the dataset. We convert these records to a ``canonical'' Parquet format, with no compression and default encoding.

\subsubsection{NYC Taxi Trip Records}
The New York City Taxi and Limousine Commission (TLC) is the agency responsible
for licensing and regulating New York City's taxis, for-hire vehicles, commuter vans, and paratransit vehicles. The TLC collects and publishes trip record
information for each taxi and for-hire vehicle trip completed~\cite{TLC}.

For our benchmark, we use Yellow and Green Taxi trip data from Q1 2025 (Jan--Mar).  We convert these records to a ``canonical'' Parquet format, with no compression and default encoding.

\subsubsection{Climate Reanalysis}
The \textit{climate reanalysis} is an important tool in climate study. The European Centre for Medium-Range Weather Forecasts (ECMWF) currently maintains ERA5, the fifth-generation of their global reanalysis dataset~\cite{ERA5}. For our benchmark, we used 5 datasets from October 1987: 10m u-component of wind (\verb|ERA5_wind|), mean sea level pressure (\verb|ERA5_pressure|), snow density (\verb|ERA5_snow|), downward UV radiation at the surface (\verb|ERA5_flux|), and total precipitation (\verb|ERA5_precip|). For each dataset, there are 720 hourly snapshots.

\subsubsection{US Census}
The anonymized 2020 US Census data is available via the \textit{Privacy-Protected Microdata File} (PPMF)~\cite{ppmf2024}. The PPMF data are organized into two huge CSV files containing household data (\texttt{ppmf\_unit}) and people data (\texttt{ppmf\_person}). We preprocessed the files by breaking them into 100 MB chunks, splitting between line breaks. This generates 548 \texttt{ppmf\_unit} files and 1,256 \texttt{ppmf\_person} files.

The Census Bureau also collects the yearly American Community Survey (ACS). The data are available via the \textit{Public Use Microdata Sample} (PUMS)~\cite{pums2023}. We build our corpus from the 5-year PUMS data from 2023~\cite{pums2023}. There are two sets of CSV files, one for people and one for households. We refer to these datasets as \texttt{psam\_p} and \texttt{psam\_h}.

\subsection{Hardware and Software}
All benchmarks were run on a Lenovo P620 desktop with an AMD Ryzen Threadripper PRO 3995WX CPU, with 256 GB of memory (8$\times$32 GB DDR4 3200MHz RDIMM ECC memory), and a 2 TB Samsung PM981a SSD. For a fair comparison, swap was not used. All benchmarks operate single-threaded in memory. Precision Boost was disabled for consistent speed numbers.

We compiled \NAME{} with GCC~14 on Fedora Linux~41. Each dataset is also benchmarked against a list of widely-used traditional compressors and some modern deep-learning systems. We chose XZ, Zstd, and gzip/zlib as representative traditional compressors; we chose cmix and NNCP to represent modern deep-learning systems. For a fair speed comparison we did not allow GPU acceleration, which limited our choice of ML-based compressors to those that supported CPU computation.

Among the traditional compressors, XZ is well-regarded for its compression ratio and Zstd for its speed. Cmix is long-known for being the high watermark for compression ratio, often at the top of generic compression benchmarks.

\subsection{Compressor Generation and Training}
\label{subsections:compressor-training}

For each dataset, we built \NAME{} compressors following the structure described in the SAO example in \Cref{sections:compressor_generation_sao}: Parse into homogeneous streams, cluster similar streams, and compress the clusters. The parsers for CSV and Parquet were written manually (the GRIB data are just numeric arrays, so no parsing was needed). Diverging from SAO, we decided to automate the clustering and backend graph generation processes using a custom-purpose training script \footnote{The automated trainer is open-source and accessible by calling \texttt{zli train}. Usage details can be found in the online \href{https://openzl.org/getting-started/cli/}{\textcolor{blue}{\underline{quick-start guide}}}.}.

The training script handles clustering and backend graph generation in separate training stages. After parsing the input data into a set of streams, as in the SAO example, the clustering trainer groups similar streams. Initially, each stream is assigned to a different ``cluster''. The trainer greedily combines pairs of clusters whose combined compressed size is smaller than the summed individual compressed sizes. It then repeats the process until it reaches a local minimum.

The backend graph generator uses the NSGA-II genetic algorithm~\cite{DEB02} to build a Pareto-optimal set of compression graphs. Each compression graph is a DAG that can be manipulated by the algorithm. The population is seeded with a set of simple but commonly effective compression graphs. The crossover and mutation functions are taken from Genetic Programming~\cite{koza92}, which is a natural fit because a compression graph is just a reversible computation graph.

Finally, the sets of Pareto-optimal backend graphs for each cluster are merged iteratively. Each set of $n$ graphs is merged into the accumulated Pareto-optimal set, then the merged set is pruned back down to $n$ entries by iteratively selecting the points with the highest crowding distance~\cite{DEB02}.

\Cref{table:test-train-split} gives some summary statistics on the training procedure for each test dataset. Anecdotally, we found that training on more data does not necessarily improve the performance of trained compressors. To use the SAO example again, we found that training on the first 1\% of the data increases compression ratio by 29\%---but training on the \emph{entire} dataset increases compression ratio by only an additional 3\%. This example suggests that performance plateaus quickly after building a representative sample. We hypothesize this is because semantic structure is high-signal so adding more training data only helps by preventing overfitting. For our experimental datasets, we used at least 3 files when training and aimed to use a test-train split of 99-1 for larger datasets.

\begin{table}[ht]
    \caption{Training on Experimental Datasets}
    \label{table:test-train-split}
    \centering
    \begin{NiceTabular}{l S S S}
    \toprule
    Dataset & {\thead{Training Set\\ Size (MiB)}} & {\thead{\% of Total\\ Dataset Size}} & {\thead{Training Speed \\ (MiB/min)}}\\
    \midrule
    \texttt{binance} & 149.3 & 15.50\% & 3.73\\
    \texttt{tlc} & 12.9 & 0.87\% & 5.83\\
    \texttt{era5\_flux} & 39.6 & 0.69\% & 7.45\\
    \texttt{era5\_precip} & 39.6 & 0.69\% & 9.69\\
    \texttt{era5\_pressure} & 39.6 & 0.69\% & 5.66\\
    \texttt{era5\_snow} & 39.6 & 0.69\% & 11.60\\
    \texttt{era5\_wind} & 39.6 & 0.69\% & 4.34\\
    \texttt{ppmf\_person} & 1334.5 & 1.10\% & 3.77\\
    \texttt{ppmf\_unit} & 953.8 & 1.83\% & 4.04\\
    \texttt{psam\_h} & 434.3 & 10.80\% & 1.54\\
    \texttt{psam\_p} & 979.7 & 9.83\% & 1.12\\
    \bottomrule  
    \end{NiceTabular}
\end{table}

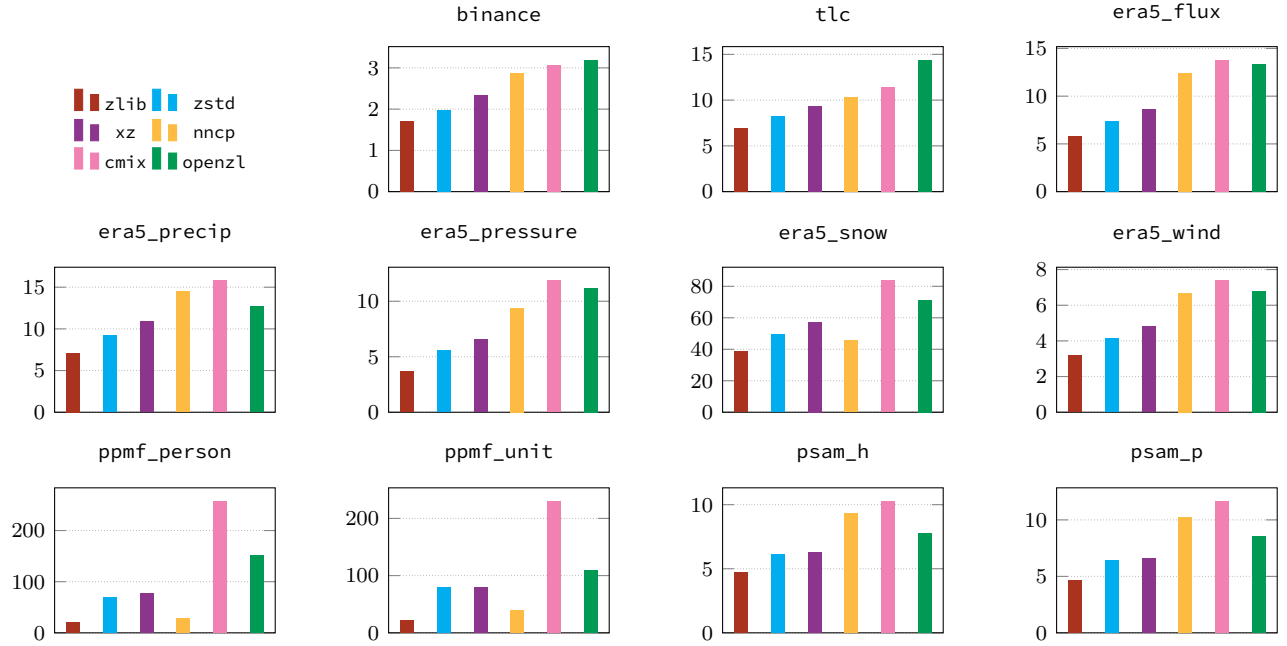
\begin{figure*}[ht!]
    \centering
\pgfplotsset{compat=1.18}
\begin{tikzpicture}
\begin{groupplot}[
  name=gp,
  group style={
    group size=4 by 3,
    horizontal sep=1.5cm,
    vertical sep=1.0cm,
  },
  width=4.5cm,
  height=3.5cm,
  ymin=0,
  ybar,
  /pgf/bar width = 5pt,
  /pgf/bar shift=0pt,
  ymajorgrids,
  major grid style={ultra thin, densely dotted, black},
  xtick style={draw=none},
  symbolic x coords={zlib,zstd,xz,nncp,cmix,openzl},
  xticklabels={},
  title style={font=\footnotesize\ttfamily},
  tick label style={font=\footnotesize}
]
\nextgroupplot[
  axis lines=none,
  ticks=none,
  title={},
  xmin=zlib, xmax=openzl, ymin=0, ymax=1,
  legend columns=2,
  legend style={draw=none,fill=none,font=\scriptsize,
    at={(0.05,0.75)},
    anchor=north west,
  },
]
\addlegendimage{ybar, rpbar, zlibplotcolor}   \addlegendentry{\tt zlib}
\addlegendimage{ybar, rpbar, zstdplotcolor}   \addlegendentry{\tt zstd}
\addlegendimage{ybar, rpbar, xzplotcolor}     \addlegendentry{\tt xz}
\addlegendimage{ybar, rpbar, nncpplotcolor}   \addlegendentry{\tt nncp}
\addlegendimage{ybar, rpbar, cmixplotcolor}   \addlegendentry{\tt cmix}
\addlegendimage{ybar, rpbar, openzlplotcolor} \addlegendentry{\tt openzl}
\nextgroupplot[title={binance}]
\addplot[rpbar,zlibplotcolor]   coordinates {(zlib,1.7062)};
\addplot[rpbar,zstdplotcolor]   coordinates {(zstd,1.9739)};
\addplot[rpbar,xzplotcolor]     coordinates {(xz,2.3321)};
\addplot[rpbar,nncpplotcolor]   coordinates {(nncp,2.8695)};
\addplot[rpbar,cmixplotcolor]   coordinates {(cmix,3.0614)};
\addplot[rpbar,openzlplotcolor] coordinates {(openzl,3.1954)};
\nextgroupplot[title={tlc}]
\addplot[rpbar,zlibplotcolor]  coordinates {(zlib,6.9881)};
\addplot[rpbar,zstdplotcolor]  coordinates {(zstd,8.2576)};
\addplot[rpbar,xzplotcolor]    coordinates {(xz,9.3371)};
\addplot[rpbar,nncpplotcolor]  coordinates {(nncp,10.3674)};
\addplot[rpbar,cmixplotcolor]  coordinates {(cmix,11.3721)};
\addplot[rpbar,openzlplotcolor]coordinates {(openzl,14.4030)};
\nextgroupplot[title={era5\_flux}]
\addplot[rpbar,zlibplotcolor]  coordinates {(zlib,5.7837)};
\addplot[rpbar,zstdplotcolor]  coordinates {(zstd,7.4040)};
\addplot[rpbar,xzplotcolor]    coordinates {(xz,8.6505)};
\addplot[rpbar,nncpplotcolor]  coordinates {(nncp,12.3996)};
\addplot[rpbar,cmixplotcolor]  coordinates {(cmix,13.8070)};
\addplot[rpbar,openzlplotcolor]coordinates {(openzl,13.3300)};
\nextgroupplot[title={era5\_precip}]
\addplot[rpbar,zlibplotcolor]  coordinates {(zlib,7.1327)};
\addplot[rpbar,zstdplotcolor]  coordinates {(zstd,9.2040)};
\addplot[rpbar,xzplotcolor]    coordinates {(xz,10.8578)};
\addplot[rpbar,nncpplotcolor]  coordinates {(nncp,14.4997)};
\addplot[rpbar,cmixplotcolor]  coordinates {(cmix,15.8141)};
\addplot[rpbar,openzlplotcolor]coordinates {(openzl,12.7000)};
\nextgroupplot[title={era5\_pressure}]
\addplot[rpbar,zlibplotcolor]  coordinates {(zlib,3.6887)};
\addplot[rpbar,zstdplotcolor]  coordinates {(zstd,5.5620)};
\addplot[rpbar,xzplotcolor]    coordinates {(xz,6.5574)};
\addplot[rpbar,nncpplotcolor]  coordinates {(nncp,9.3988)};
\addplot[rpbar,cmixplotcolor]  coordinates {(cmix,11.8771)};
\addplot[rpbar,openzlplotcolor]coordinates {(openzl,11.2100)};
\nextgroupplot[title={era5\_snow}]
\addplot[rpbar,zlibplotcolor]  coordinates {(zlib,38.6100)};
\addplot[rpbar,zstdplotcolor]  coordinates {(zstd,49.4200)};
\addplot[rpbar,xzplotcolor]    coordinates {(xz,57.4713)};
\addplot[rpbar,nncpplotcolor]  coordinates {(nncp,45.4382)};
\addplot[rpbar,cmixplotcolor]  coordinates {(cmix,83.7434)};
\addplot[rpbar,openzlplotcolor]coordinates {(openzl,71.1900)};
\nextgroupplot[title={era5\_wind}]
\addplot[rpbar,zlibplotcolor]  coordinates {(zlib,3.1686)};
\addplot[rpbar,zstdplotcolor]  coordinates {(zstd,4.1420)};
\addplot[rpbar,xzplotcolor]    coordinates {(xz,4.8054)};
\addplot[rpbar,nncpplotcolor]  coordinates {(nncp,6.6937)};
\addplot[rpbar,cmixplotcolor]  coordinates {(cmix,7.3922)};
\addplot[rpbar,openzlplotcolor]coordinates {(openzl,6.7900)};
\nextgroupplot[title={ppmf\_person}]
\addplot[rpbar,zlibplotcolor]  coordinates {(zlib,21.3220)};
\addplot[rpbar,zstdplotcolor]  coordinates {(zstd,68.5300)};
\addplot[rpbar,xzplotcolor]    coordinates {(xz,76.9231)};
\addplot[rpbar,nncpplotcolor]  coordinates {(nncp,28.4815)};
\addplot[rpbar,cmixplotcolor]  coordinates {(cmix,257.7720)};
\addplot[rpbar,openzlplotcolor]coordinates {(openzl,150.8300)};
\nextgroupplot[title={ppmf\_unit}]
\addplot[rpbar,zlibplotcolor]  coordinates {(zlib,22.2222)};
\addplot[rpbar,zstdplotcolor]  coordinates {(zstd,79.6600)};
\addplot[rpbar,xzplotcolor]    coordinates {(xz,80.0000)};
\addplot[rpbar,nncpplotcolor]  coordinates {(nncp,39.5534)};
\addplot[rpbar,cmixplotcolor]  coordinates {(cmix,230.3319)};
\addplot[rpbar,openzlplotcolor]coordinates {(openzl,109.6800)};
\nextgroupplot[title={psam\_h}]
\addplot[rpbar,zlibplotcolor]  coordinates {(zlib,4.6917)};
\addplot[rpbar,zstdplotcolor]  coordinates {(zstd,6.1480)};
\addplot[rpbar,xzplotcolor]    coordinates {(xz,6.2518)};
\addplot[rpbar,nncpplotcolor]  coordinates {(nncp,9.2923)};
\addplot[rpbar,cmixplotcolor]  coordinates {(cmix,10.2870)};
\addplot[rpbar,openzlplotcolor]coordinates {(openzl,7.8000)};
\nextgroupplot[title={psam\_p}]
\addplot[rpbar,zlibplotcolor]  coordinates {(zlib,4.6533)};
\addplot[rpbar,zstdplotcolor]  coordinates {(zstd,6.4320)};
\addplot[rpbar,xzplotcolor]    coordinates {(xz,6.6269)};
\addplot[rpbar,nncpplotcolor]  coordinates {(nncp,10.2578)};
\addplot[rpbar,cmixplotcolor]  coordinates {(cmix,11.6719)};
\addplot[rpbar,openzlplotcolor]coordinates {(openzl,8.5200)};
\end{groupplot}
\end{tikzpicture}

    \caption{Compression ratios of competitor systems relative to \NAME{}. Higher is better.}
    \label{fig:ratio-bar-chart}
\end{figure*}

\section{Experimental Results}
\label{sections:results}

Trained \NAME{} compressors are able to beat competitors on at least one axis of performance, and sometimes all three. We start with \cref{subsections:ratio-results}, conducting a pure compression ratio comparison without considering speed. This experiment showcases the theoretical high watermark of each compressor being tested. We follow this with an analysis of the speed vs. ratio tradeoff for compressors that have this configurability. \Cref{subsections:pareto-frontier} does not compare NNCP or cmix, as they are not configurable.

\subsection{Best Compression Ratio}
\label{subsections:ratio-results}

The results of the high-watermark experiment is presented in \cref{fig:ratio-bar-chart}. Since this experiment includes cmix and NNCP, we compared performance on excerpts of each dataset. Cmix especially is too slow to process a multi-gigabyte dataset in a reasonable timeframe. On all datasets, \NAME{} compressors are able to exceed the compression ratios offered by traditional compressors and remain competitive with deep-learning compressors.


On numeric GRIB datasets, \NAME{} performs well, beating NNCP on the majority of datasets and approaching cmix on as many. Unsurprisingly, there is a lot of structure in climate data that can be exploited to improve compression. Particularly, \NAME{}'s ability to work with numeric data types sets it apart from traditional compressors, which must work byte by byte.

On the CSV datasets, \NAME{} performs less well, and markedly worse than cmix. Recall from \Cref{subsections:compressor-training} that for tabular formats, we trained compressors to cluster based on inter-column correlation. This limits our ability to exploit inter-row correlation and local correlations that don't extend to the entire file. Nonetheless, the data show that inter-column correlation is still a powerful weapon; we still perform better than the traditional LZ-based compressors. Beyond this, CSV is fundamentally a plaintext format. This means we lose the edge from working with numeric fields in GRIB.

On Parquet datasets, we can combine our strengths of semantic understanding of integers with the column-based approach. \NAME{} is able to capture the inter-column correlation by clustering similar columns, and is then able to build a specialized compression graph for each cluster of columns. Unsurprisingly, \NAME{} achieves superior performance on these datasets, beating both NNCP and cmix.

\begin{table}[ht]
    \caption{Corresponding Average Speeds (in MiB/s) for \Cref{fig:ratio-bar-chart}}
    \label{table:avg-speed}
    \centering
    \begin{NiceTabular}{ l S S }
    \CodeBefore
    \rectanglecolor{metabg}{7-1}{7-3}
    \Body
    \toprule
    Compressor & {Mean C. Speed} & {Mean D. Speed}\\
    \midrule
    \texttt{zlib -6} & 52.5 & 715.\\
    \texttt{zstd -19} & 6.07 & 2820\\
    \texttt{xz -9} & 6.14 & 314.\\
    \texttt{nncp} & 0.00252 & 0.00253\\
    \texttt{cmix} & 0.000972 & 0.000972\\
    \texttt{openzl} & 142. & 323.\\
    \bottomrule
    \end{NiceTabular}
\end{table}

While this experiment is focused on maximizing ratio, the full story would be incomplete without mentioning the difference in processing speeds between \NAME{} and its competitors. \Cref{table:avg-speed} summarizes these speed numbers. NNCP and cmix are irrecoverably slow, with both compression and decompression speeds 100,000$\times$ worse than \NAME{}. In fact, \NAME{} has the highest average compression speed out of \emph{all} compressors, and is equivalent to XZ in decompression. 


\subsection{Compressor Tradeoff Selection}
\label{subsections:pareto-frontier}

\begin{figure*}[ht!]
    \centering
    \scalebox{.7}{

\pgfplotsset{
    openzlplotstyle/.style={
        mark=openzlplotmark,
        color=openzlplotcolor
    },
    zstdplotstyle/.style={
        mark=zstdplotmark,
        color=zstdplotcolor
    },
    xzplotstyle/.style={
        mark=xzplotmark,
        color=xzplotcolor
    },
    zlibplotstyle/.style={
        mark=zlibplotmark,
        color=zlibplotcolor
    },
    bloscplotstyle/.style={
        mark=bloscplotmark,
        color=bloscplotcolor
    },
    parquetplotstyle/.style={
        mark=parquetplotmark,
        color=parquetplotcolor
    },
    algo and cpareto filter/.style n args={2}{
        x filter/.code={%
            \edef\tmpalgo{\thisrow{algorithm}}%
            \edef\tmpalgowanted{#1}%
            \edef\tmpcopt{\thisrow{in_algo_cspeed_frontier}}%
            \edef\tmpdopt{\thisrow{in_algo_dspeed_frontier}}%
            \edef\tmpcoptwanted{#2}%
            \ifx\tmpalgo\tmpalgowanted
                \ifx\tmpcopt\tmpcoptwanted
                \else
                    \def\pgfmathresult{}
                \fi
            \else
                \def\pgfmathresult{}
            \fi
        },
    },
    algo and dpareto filter/.style n args={2}{
        x filter/.code={%
            \edef\tmpalgo{\thisrow{algorithm}}%
            \edef\tmpalgowanted{#1}%
            \edef\tmpcopt{\thisrow{in_algo_cspeed_frontier}}%
            \edef\tmpdopt{\thisrow{in_algo_dspeed_frontier}}%
            \edef\tmpdoptwanted{#2}%
            \ifx\tmpalgo\tmpalgowanted
                \ifx\tmpdopt\tmpdoptwanted
                \else
                    \def\pgfmathresult{}
                \fi
            \else
                \def\pgfmathresult{}
            \fi
        },
    },
    speed chart/.style={
        width=6cm,
        height=5cm,
        x unit=\si{\mebi\byte\per\second},
        y unit=\si{\byte\per\byte},
        log ticks with fixed point,
        xmajorgrids,
        ymajorgrids,
        major grid style={ultra thin, densely dotted, black},
        log basis y=2,
        yminorticks=true,
        legend columns=-1,
        legend style={draw=none, fill=none,font=\scriptsize, at={(-1.05,1.4)},
            anchor=south west,
        },
        label style={
            font=\footnotesize,
        },
    },
    cspeed chart/.style={
        speed chart,
        xlabel={Compression Speed},
        ylabel={Compression Ratio},
        xshift=-.15cm,
        anchor=east,
    },
    dspeed chart/.style={
        speed chart,
        xlabel={Decompression Speed},
        y unit={},
        yticklabel pos=right,
        xshift=.15cm,
        anchor=west,
    },
    speed chart legend/.style={},
}

\begin{tikzpicture}[global]

\coordinate (p1) at (-5.625,   0);
\coordinate (p2) at ( 5.625,   0);
\coordinate (p3) at (-5.625,  -5.5);
\coordinate (p4) at ( 5.625,  -5.5);
\coordinate (p5) at (-5.625, -11);
\coordinate (p6) at ( 5.625, -11);
\coordinate (p7) at ( 0.0  , -16.5);

\def\tmpplots{}%

\edef\legendplot{(p1)}%

\pgfplotsforeachungrouped \tablename / \plotpos in {
    {result_tables/pareto_frontiers/binance_new.csv}       /(p1),
    {result_tables/pareto_frontiers/tlc_new.csv}           /(p2),
    {result_tables/pareto_frontiers/era5_flux_new.csv}     /(p3),
    {result_tables/pareto_frontiers/era5_precip_new.csv}   /(p4),
    {result_tables/pareto_frontiers/ppmf_unit.csv}         /(p5),
    {result_tables/pareto_frontiers/psam_p_new.csv}        /(p6)%
} {

    \eappto\tmpplots{
        \noexpand\begin{semilogxaxis}[
            cspeed chart,
            at=\plotpos
        ]
    }

    \pgfplotsforeachungrouped \algorithm / \algostyle in {
        openzl/openzlplotstyle,
        zstd/zstdplotstyle,
        xz/xzplotstyle,
        zlib/zlibplotstyle
    } {
        \eappto\tmpplots{
            \noexpand\addplot [
                \algostyle,
                mark options={
                    scale=0.75
                },
                algo and cpareto filter={\algorithm}{True},
            ] table [
                x expr={\noexpand\thisrow{compress_speed_mbps}},
                y expr={\noexpand\thisrow{compression_ratio}},
                col sep=comma,
                row sep=newline,
            ] {\tablename};

            \noexpand\addplot [
                \algostyle,
                only marks,
                forget plot,
                mark options={
                    scale=0.75
                },
                algo and cpareto filter={\algorithm}{False},
            ] table [
                x expr={\noexpand\thisrow{compress_speed_mbps}},
                y expr={\noexpand\thisrow{compression_ratio}},
                col sep=comma,
                row sep=newline,
            ] {\tablename};
        }
    }

    \eappto\tmpplots{
        \noexpand\end{semilogxaxis}
    }

    \eappto\tmpplots{
        \noexpand\begin{semilogxaxis}[
            dspeed chart,
            \ifx\plotpos\legendplot%
                speed chart legend,
            \fi%
            at=\plotpos
        ]
    }

    \pgfplotsforeachungrouped \algorithm / \algostyle in {
        openzl/openzlplotstyle,
        zstd/zstdplotstyle,
        xz/xzplotstyle,
        zlib/zlibplotstyle
    } {
        \eappto\tmpplots{
            \noexpand\addplot [
                \algostyle,
                mark options={
                    scale=0.75
                },
                algo and dpareto filter={\algorithm}{True},
            ] table [
                x expr={\noexpand\thisrow{decompress_speed_mbps}},
                y expr={\noexpand\thisrow{compression_ratio}},
                col sep=comma,
                row sep=newline,
            ] {\tablename};

            \noexpand\addplot [
                \algostyle,
                only marks,
                forget plot,
                mark options={
                    scale=0.75
                },
                algo and dpareto filter={\algorithm}{False},
            ] table [
                x expr={\noexpand\thisrow{decompress_speed_mbps}},
                y expr={\noexpand\thisrow{compression_ratio}},
                col sep=comma,
                row sep=newline,
            ] {\tablename};


            \ifx\plotpos\legendplot%
                \noexpand\addlegendentry{\noexpand\tt\noexpand\footnotesize\algorithm};
            \fi
        }
    }

    \eappto\tmpplots{
        \noexpand\end{semilogxaxis}
    }

}

\tmpplots

\node [txt] at ($(p1)+(0,2.25)$) {binance};
\node [txt] at ($(p2)+(0,2.25)$) {tlc};
\node [txt] at ($(p3)+(0,2.25)$) {era5\_flux};
\node [txt] at ($(p4)+(0,2.25)$) {era5\_precip};
\node [txt] at ($(p5)+(0,2.25)$) {ppmf\_unit};
\node [txt] at ($(p6)+(0,2.25)$) {psam\_p};

\end{tikzpicture}
}
    
    \caption{Compression and decompression Pareto frontiers of different algorithms for selected datasets.}
    \label{figure:pareto_frontiers}
\end{figure*}
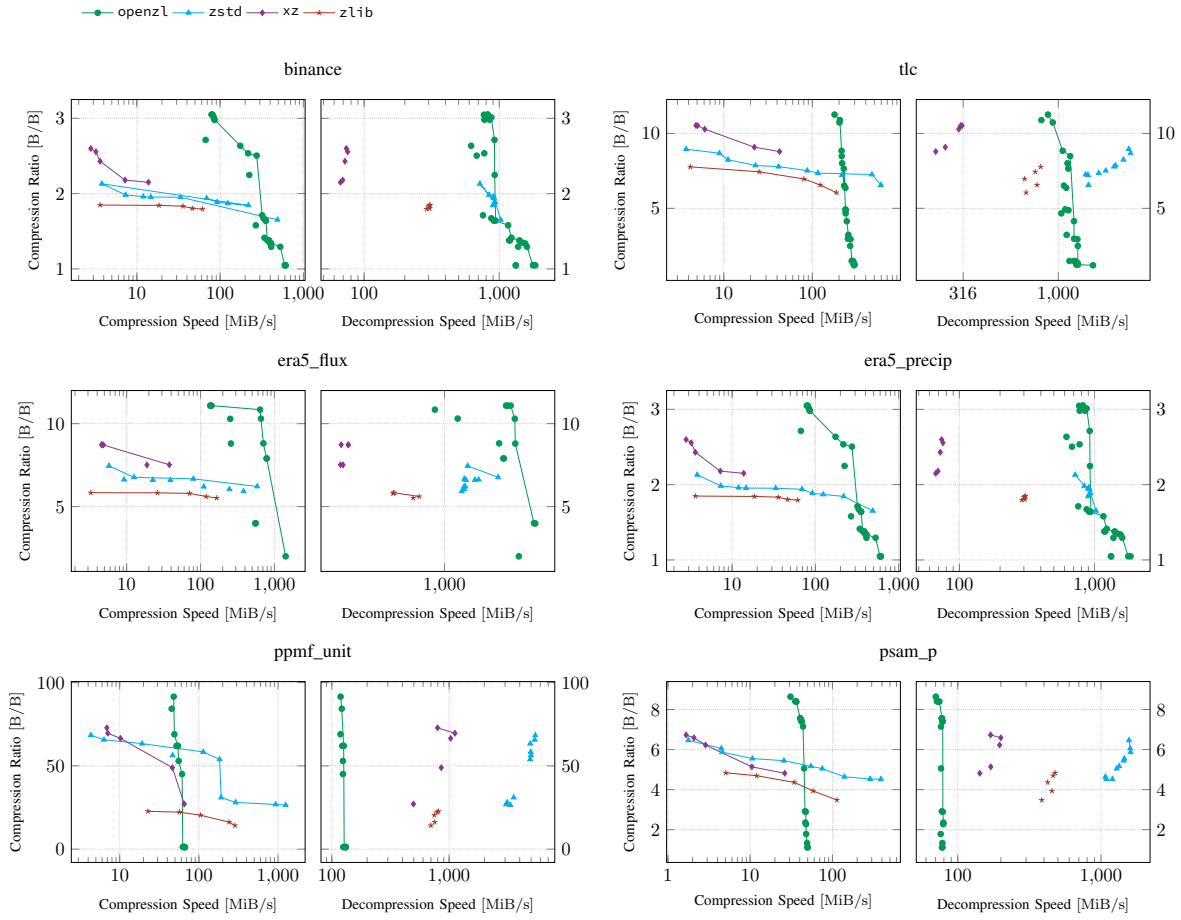

\NAME{} is not limited to pursuing aggressive compression ratios, and indeed neither are most production compressors. In addition to the ratio-focused compressors trained in the previous section, the custom trainer generates a Pareto-optimal frontier of compression graphs. We benchmarked these tradeoff points against the typical level system featured by other generic compressors. This experiment mirrors the analysis a production engineer would do when choosing a compressor to deploy. Often, speed is just as important as ratio; on compute-bound workloads, it may even be more important.

Every point on the plot represents a unique compression graph for \NAME{}, or a unique compression level for other compressors. In many cases, the \NAME{} tradeoff curve for ratio vs.\; compression speed strictly dominates. In particular, on the Parquet and GRIB datasets, it would \emph{never} be better to choose XZ or zlib, because there exists an \NAME{} compressor that pareto-dominates it. However, for the CSV datasets \NAME{} first has to parse the CSV file, which limits its speed.

Here, the power of the graph model is fully realized. Traditional compressors are limited to one pipeline of operations, or a small handful of predefined pipelines, so the performance tradeoffs they can offer are limited. Since \NAME{} can create an entirely new compression graph for each point in its Pareto-optimal frontier, it is able to offer a significantly wider range of tradeoffs along all axes. 

Of course, this power comes with a higher setup cost than a generic compressor, which only requires selecting a compression level. Without an effective parser and graph (automatically generated or otherwise), \NAME{} won't perform any better than traditional compressors.

\section{\NAME{} Deployment at Meta}
\label{sections:openzl_at_meta}

Optimizing compression performance tradeoffs is important for enterprises operating at Meta's scale. Prior to \NAME{}, Zstd was the primary compressor used at Meta, since it offered Pareto-optimal performance across a wide variety of use cases~\cite{JEONG22}. This was the result of a more than decade-long drive for compression efficiency, achieved both by optimizing Zstd's performance and converting uses of compression to Zstd. It eventually became clear that the headroom for further improvements from Zstd, within Zstd, or even with LZ compression in general, was fundamentally limited. Thus \NAME{} was born.

\NAME{} has now replaced a meaningful fraction of Zstd use in production at Meta. Much of the data at Meta is serialized using Thrift~\cite{thrift}, both at rest and in transit. A custom parser that understands Thrift, in conjunction with training tools to specialize the compressor for individual use cases, allows \NAME{} to effectively compress a wide range of traffic.

\NAME{} integrations at Meta can be split into 3 broad and interconnected categories:

\begin{enumerate}[label=(\roman*)]
    \item \textbf{Data warehouses:}
    Data Warehouses store vast amounts of data for analytics and training, typically stored in a columnar databases. Compression in columnar databases is not a new technique; open-source formats like Parquet have already explored this to some success. However, the expressibility allowed by the graph model allows \NAME{} to extract additional gains by better modeling the data.
    \item \textbf{Training data:}
    Training data encompasses everything that models are trained on. It flows from the applications where it is logged, through the online and offline training pipelines, and to the training jobs. The same data is also stored and cached for inference. At each step along the way, the data is compressed to improve efficiency.
    \item \textbf{Model training:}
    In addition to the training data, the models themselves are also compressed. During training, model checkpoints are saved frequently, so that training progress isn't lost if the job fails. Then, when a new model is shipped, that checkpoint must be saved through the model lifetime. Compressing these models reduces storage, checkpoint overhead, and distribution bandwidth.
\end{enumerate}

\begin{table}[h!]
    \caption{An overview of major \NAME{} integrations at Meta, as of the time of writing}
    \label{table:openzl_at_meta}
    \centering
    \begin{NiceTabular}{lcccccc}
    \CodeBefore
    \Body
    \toprule
    Project                   & Use Case          & Data Format                & Trained\\
    \midrule
    Nimble                    & Data warehouse    & Raw Columns                & No\\
    Scribe                    & Data warehouse    & Thrift                     & Yes\\
    Feature storage           & Training data     & Thrift                     & Yes\\
    Log aggregator            & Training data     & Thrift                     & Yes\\
    Embedding storage         & Training data     & Uncompressed \texttt{.zip} & No\\
    \makecell{PyTorch model \\ checkpoints} & Model training    & Float arrays               & No\\
    \bottomrule
    \end{NiceTabular}
\end{table}

\Cref{table:openzl_at_meta} provides a summary table of internal \NAME{} deployments so far, with details for each service listed below.

\begin{description}
    \item[Nimble:] Nimble \cite{NIMBLE} is a columnar database format used at Meta. Previously, Zstd was the only backend compressor offered by Nimble. Replacing Zstd with (untrained) \NAME{} compressors immediately saved 10\% compressed size compared to this baseline. Most of these gains came from labeling numeric data types as such, and stacks on top of the transformations that Nimble uses to pre-process its data\footnote{Nimble applies transformations that improve query efficiency by operating on the encoded data, where \NAME{} applies transformations that are useful for compression only.}. Traditional compressors like Zstd compress bytes, so the columnar formats serialize numeric data to bytes before compressing, typically by Varint encoding. Since \NAME{} operates directly on numeric data, it is able to skip this step, which improves query efficiency.
    
    \item[Scribe:] Scribe \cite{SCRIBE} is a log processing system used extensively at Meta. With training, \NAME{} improved compression ratios by $\sim$15\% compared to Zstd. This improvement is felt both in reduced storage costs and in higher network throughput. Increasing throughput has the additional benefit of improving training data quality since fewer records are dropped during traffic peaks. As noted in \cite{SCRIBE}, Scribe data is constantly evolving. We've been able to maintain these ratio wins via regular training runs similar to the ones ran for the benchmark experiments.
    
    \item[PyTorch model checkpoints:] Model checkpoints are saved frequently during the training process. Ephemeral checkpoints are only stored for a short time, but anchor checkpoints are saved for the model's lifetime. \NAME{} leverages type information to save an average of 17\% on storage for model checkpoints by compressing the floating point exponents, with savings varying based on the floating point data type. Checkpoint saving and loading is network bound, so the 17\% checkpoint size reduction also reduces checkpoint overhead by 15\% on training machines, which improves GPU utilization.
    
    \item[Feature storage:] This service caches features stored in Thrift format for training data generation. Similar to Scribe, \NAME{} was deployed with training for Thrift data. However, the Feature storage team chose a different trade-off point on the speed-ratio curve. Switching to \NAME{} reduced storage by 10\% and CPU utilization by 5\% compared to Zstd level 6. Moreover, this was done solely by reusing existing components already built for Scribe.
    
    \item[Log aggregator:] This service joins realtime logs from several data sources for training data generation. \NAME{} was able to reduce compressed size by 18\% compared to Zstd. Initially, \NAME{} was configured to compress the Thrift data directly. But, due to the latency sensitive nature of the service, they migrated to directly compressing arrays of integers with \NAME{}. This allows them to skip the Thrift serialization \& deserialization stages.
    
    \item[Embedding storage:] Compressing \texttt{bfloat16} embeddings serialized in  PyTorch's \texttt{torch.save()} format reduces compressed size by 30\% (thus allowing us to store 43\% more training data). Prior to \NAME{}, compression was not deemed computationally profitable because traditional compressors struggle with floating-point data. Zstd, for instance, can't shrink the data by more than $\sim$10\%, even at the highest compression level. By generating a floating-point compressor, we were able to save storage while not regressing training performance. Beyond this, the development timeline for this new compressor was on the order of days due to reuse of existing components developed for PyTorch model checkpoints.
\end{description}

Overall, OpenZL has helped Meta bend the curve of AI growth. The compression improvements that OpenZL offers allow Meta to do more with the same amount of hardware---better training data compression means more data can be pushed through the same pipe; smaller data means less compute is spent reading data from the data warehouse; reduced network traffic for model checkpointing improves GPU utilization.

\subsection{Training with Managed Compression}
\label{subsections:training}

\NAME{}'s modular treatment of the components of compression, and the development of tools that automate the configuration and composition of those components, mean that \NAME{} lends itself well to offline training as described in \Cref{subsections:compressor-training}. Although the configurations under consideration internally are different and more diverse, this training workflow nonetheless has the same overall shape as training a dictionary for Zstd.

\begin{figure}[h]
    \centering
    \begin{tikzpicture}
        \draw [dotted] (-3.75, 0) -- (3.25, 0);
        \node [txt, tiny, anchor=south west] at (-3.75, 0) {online};
        \node [txt, tiny, anchor=north west] at (-3.75, 0) {offline};
        
        \node [txt] at ( 0,  1.75) (MC_LIB) {Managed\\Compression\\Library};
        
        \node [ss, txt, above=0.375 of MC_LIB.north, anchor=south] (USER) {Users};
        
        \node [fn, txt, below=0.375 of MC_LIB.south, anchor=north] (OZL) {OpenZL};
        
        \node [fn, txt, fill=white] at (2.5, 0) (DATASTORE) {Data\\Store};
        
        \node [txt] at ( 0, -1.75) (TRAINER) {Managed\\Compression\\Automation};
        
        \node [fn, txt, below=0.375 of TRAINER.south, anchor=north] (OZL_TRAINER) {OpenZL Trainer};
        
        \node [fn, txt, fill=white] at (-2.5,  0) (CFGSTORE)  {Config\\Store};
        
        \draw [<->]
              (USER)
              to
              (MC_LIB);
        \draw [<->]
              (MC_LIB)
              to [out=270, in=90]
              (OZL);
        \draw [->]
              (MC_LIB)
              to [out=  0, in= 90]
              node [midway, above, sloped, tiny] {samples}
              (DATASTORE);
        \draw [<->]
              (TRAINER)
              to [out=270, in=90]
              (OZL_TRAINER);
        \draw [->]
              (DATASTORE)
              to [out=270, in=  0]
              (TRAINER);
        \draw [->]
              (TRAINER)
              to [out=180, in=270]
              node [tiny, txt, below, sloped, pos=.5] {configs}
              (CFGSTORE);
        \draw [->]
              (CFGSTORE)
              to [out= 90, in=180]
              (MC_LIB);
    \end{tikzpicture}
    \caption{\NAME{} integrated into Managed Compression.}
\end{figure}
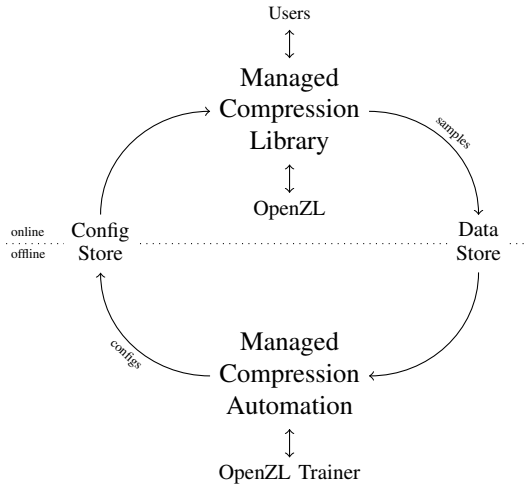

And in fact, Meta's Managed Compression system~\cite{HANDTE18}, which was originally designed to manage dictionaries for Zstd compression, has proven adept at training compressors for \NAME{}. The trainer accepts a corpus of representative samples and an existing compressor; it can configure and parameterize nodes or even construct and replace sub-graphs throughout the compressor. After validation and benchmarking, the resulting compressor can then be re-serialized and deployed to the fleet.

This architecture has proven useful not only to find useful \NAME{} configurations, but also to drive broad adoption of \NAME{} at Meta: in the same way that this systematic approach to training made Zstd dictionary adoption frictionless, this infrastructure makes \NAME{} easy to integrate, and thousands of trained \NAME{} compressors are deployed to use cases at Meta through Managed Compression.

\section{Conclusions}
\label{sections:conclusions}

In this paper, we propose a graph model of compression. This new theoretical model encourages composition of small and simple codecs. We demonstrate its effectiveness with \NAME{}, a robust implementation of the graph model. The result is a scalable, production-ready system that offers unprecedented performance across diverse datasets. For a wide range of benchmark datasets, \NAME{} is able to beat the best compression ratio offered by \texttt{xz} at an order of magnitude faster compression, despite needing to parse and understand the data. \NAME{} demonstrates that the compression efficiency unlocked by understanding the data far outweighs the effort spent on parsing, and that this can be achieved via composition of a library of relatively simple codecs.

We hope that the positive results from Meta-internal use cases will motivate data owners to investigate their own wins from using \NAME{}. In particular, we expect the automated training tools presented to be more than adequate to achieve compression wins that justify the resource investment in \NAME{}.

\subsection{Future Work}

The unreasonable effectiveness of our first foray into training leads us to believe that the graph model is uniquely positioned to facilitate ML-guided generation of compressors. We are tempted to view this as ``the next big thing'' in production-scale compression. Whereas compression research has up to now eluded those without domain expertise, we believe the future of application-specific compressors will be unlocked via investment in automated learning methods.


\section*{Acknowledgement}

We would like to thank Graham Cormode for his guidance through the publication process, especially with regard to narrative clarity and prioritization. We would like to thank Evan West for his input and advice on the text of the paper. We would like to thank our former interns Timothy Oei, Pedro Valero, Aryan Gandevia, and Faizaan Baig for their contributions to \NAME{}. Finally, we would like to thank Aras Pranckevi\v{c}ius and Takayuki Matsuoka for beta-testing \NAME{} before its open-source launch.

\IEEEtriggeratref{57}
\IEEEtriggercmd{\enlargethispage{-4.85in}}

\bibliographystyle{IEEEtran}
\bibliography{IEEEabrv,paper}

\begin{thebibliography}{10}
\providecommand{\url}[1]{#1}
\csname url@samestyle\endcsname
\providecommand{\newblock}{\relax}
\providecommand{\bibinfo}[2]{#2}
\providecommand{\BIBentrySTDinterwordspacing}{\spaceskip=0pt\relax}
\providecommand{\BIBentryALTinterwordstretchfactor}{4}
\providecommand{\BIBentryALTinterwordspacing}{\spaceskip=\fontdimen2\font plus
\BIBentryALTinterwordstretchfactor\fontdimen3\font minus
  \fontdimen4\font\relax}
\providecommand{\BIBforeignlanguage}[2]{{%
\expandafter\ifx\csname l@#1\endcsname\relax
\typeout{** WARNING: IEEEtran.bst: No hyphenation pattern has been}%
\typeout{** loaded for the language `#1'. Using the pattern for}%
\typeout{** the default language instead.}%
\else
\language=\csname l@#1\endcsname
\fi
#2}}
\providecommand{\BIBdecl}{\relax}
\BIBdecl

\bibitem{KNOLL12}
B.~Knoll and N.~d. Freitas, ``A machine learning perspective on predictive
  coding with paq8,'' in \emph{2012 Data Compression Conference}, April 2012,
  pp. 377--386.

\bibitem{CMIX}
B.~Knoll, ``Cmix compressor,'' \url{https://www.byronknoll.com/cmix.html},
  2014, accessed: 2025-04-30.

\bibitem{NNCP}
\BIBentryALTinterwordspacing
F.~Bellard, ``Lossless data compression with neural networks,'' 2019. [Online].
  Available: \url{https://bellard.org/nncp/nncp.pdf}
\BIBentrySTDinterwordspacing

\bibitem{GOYAL20}
M.~Goyal, K.~Tatwawadi, S.~Chandak, and I.~Ochoa, ``Dzip: Improved
  general-purpose lossless compression based on novel neural network
  modeling,'' in \emph{2020 Data Compression Conference (DCC)}, March 2020, pp.
  372--372.

\bibitem{LSTM-COMPRESS}
B.~Knoll, ``lstm-compress,'' \url{https://github.com/byronknoll/lstm-compress},
  2017.

\bibitem{MAO22}
\BIBentryALTinterwordspacing
Y.~Mao, Y.~Cui, T.-W. Kuo, and C.~J. Xue, ``Trace: A fast transformer-based
  general-purpose lossless compressor,'' in \emph{Proceedings of the ACM Web
  Conference 2022}, ser. WWW '22.\hskip 1em plus 0.5em minus 0.4em\relax New
  York, NY, USA: Association for Computing Machinery, 2022, p. 1829–1838.
  [Online]. Available: \url{https://doi.org/10.1145/3485447.3511987}
\BIBentrySTDinterwordspacing

\bibitem{LIU22}
\BIBentryALTinterwordspacing
A.~Liu, S.~Mandt, and G.~V. den Broeck, ``Lossless compression with
  probabilistic circuits,'' in \emph{International Conference on Learning
  Representations}, 2022. [Online]. Available:
  \url{https://openreview.net/forum?id=X_hByk2-5je}
\BIBentrySTDinterwordspacing

\bibitem{ZHANG24}
\BIBentryALTinterwordspacing
B.~Zhang, D.~Cheng, Y.~Zhang, F.~Liu, and W.~Chen, ``Compression for better: A
  general and stable lossless compression framework,'' 2024. [Online].
  Available: \url{https://arxiv.org/abs/2412.06868}
\BIBentrySTDinterwordspacing

\bibitem{GUPTA17}
A.~Gupta, A.~Bansal, and V.~Khanduja, ``Modern lossless compression techniques:
  Review, comparison and analysis,'' in \emph{2017 Second International
  Conference on Electrical, Computer and Communication Technologies
  (ICECCT)}.\hskip 1em plus 0.5em minus 0.4em\relax IEEE, 2017, pp. 1--8.

\bibitem{zstandard}
\BIBentryALTinterwordspacing
Y.~Collet, ``Zstandard - fast real-time compression algorithm,'' Facebook, Open
  source project, 2015. [Online]. Available:
  \url{https://github.com/facebook/zstd}
\BIBentrySTDinterwordspacing

\bibitem{JUMAR18}
\BIBentryALTinterwordspacing
R.~Jumar, H.~Maaß, and V.~Hagenmeyer, ``Comparison of lossless compression
  schemes for high rate electrical grid time series for smart grid monitoring
  and analysis,'' \emph{Computers \& Electrical Engineering}, vol.~71, pp.
  465--476, 2018. [Online]. Available:
  \url{https://www.sciencedirect.com/science/article/pii/S0045790617334791}
\BIBentrySTDinterwordspacing

\bibitem{THEVENIN22}
\BIBentryALTinterwordspacing
M.~Thevenin, S.~Pigoury, O.~Thomine, and F.~Gouillon, ``A comparison of
  lossless compression algorithms for altimeter data,'' \emph{EGUsphere}, vol.
  2022, pp. 1--28, 2022. [Online]. Available:
  \url{https://egusphere.copernicus.org/preprints/2022/egusphere-2022-1094/}
\BIBentrySTDinterwordspacing

\bibitem{IQBAL20}
K.~Iqbal, N.~Khan, and M.~G. Martini, ``Performance comparison of lossless
  compression strategies for dynamic vision sensor data,'' in \emph{ICASSP 2020
  - 2020 IEEE International Conference on Acoustics, Speech and Signal
  Processing (ICASSP)}, May 2020, pp. 4427--4431.

\bibitem{2BrightSparks24}
\BIBentryALTinterwordspacing
Jan 2024. [Online]. Available:
  \url{https://help.2brightsparks.com/support/solutions/articles/43000335985-comparison-of-compression-methods-and-levels}
\BIBentrySTDinterwordspacing

\bibitem{LinuxReviews}
\BIBentryALTinterwordspacing
2019. [Online]. Available:
  \url{https://linuxreviews.org/Comparison_of_Compression_Algorithms}
\BIBentrySTDinterwordspacing

\bibitem{PIBIRI22}
\BIBentryALTinterwordspacing
G.~E. Pibiri, ``Sparse and skew hashing of k-mers,'' \emph{Bioinformatics},
  vol.~38, no. Supplement\_1, pp. i185--i194, 06 2022. [Online]. Available:
  \url{https://doi.org/10.1093/bioinformatics/btac245}
\BIBentrySTDinterwordspacing

\bibitem{PIBIRI23}
\BIBentryALTinterwordspacing
------, ``On weighted k-mer dictionaries,'' \emph{Algorithms for Molecular
  Biology}, vol.~18, no.~1, p.~3, 2023. [Online]. Available:
  \url{https://doi.org/10.1186/s13015-023-00226-2}
\BIBentrySTDinterwordspacing

\bibitem{ALANKO25}
\BIBentryALTinterwordspacing
J.~N. Alanko, E.~Biagi, J.~Mackenzie, and S.~J. Puglisi, ``Batched k-mer lookup
  on the spectral burrows-wheeler transform,'' in \emph{2025 Proceedings of the
  Symposium on Algorithm Engineering and Experiments (ALENEX)}, 2025, pp.
  95--106. [Online]. Available:
  \url{https://epubs.siam.org/doi/abs/10.1137/1.9781611978339.8}
\BIBentrySTDinterwordspacing

\bibitem{CHANDAK18}
\BIBentryALTinterwordspacing
S.~Chandak, K.~Tatwawadi, I.~Ochoa, M.~Hernaez, and T.~Weissman, ``Spring: a
  next-generation compressor for fastq data,'' \emph{Bioinformatics}, vol.~35,
  no.~15, pp. 2674--2676, 12 2018. [Online]. Available:
  \url{https://doi.org/10.1093/bioinformatics/bty1015}
\BIBentrySTDinterwordspacing

\bibitem{LAN21}
\BIBentryALTinterwordspacing
D.~Lan, R.~Tobler, Y.~Souilmi, and B.~Llamas, ``Genozip: a universal extensible
  genomic data compressor,'' \emph{Bioinformatics}, vol.~37, no.~16, pp.
  2225--2230, 02 2021. [Online]. Available:
  \url{https://doi.org/10.1093/bioinformatics/btab102}
\BIBentrySTDinterwordspacing

\bibitem{CHANDRA12}
P.~Chanda, E.~Elhaik, and J.~S. Bader, ``Hapzipper: sharing hapmap populations
  just got easier,'' \emph{Nucleic acids research}, vol.~40, no.~20, pp.
  e159--e159, 2012.

\bibitem{MENTZER20}
F.~Mentzer, L.~Van~Gool, and M.~Tschannen, ``Learning better lossless
  compression using lossy compression,'' in \emph{2020 IEEE/CVF Conference on
  Computer Vision and Pattern Recognition (CVPR)}, June 2020, pp. 6637--6646.

\bibitem{TAUBMAN19}
D.~Taubman, A.~Naman, R.~Mathew, M.~Smith, O.~Watanabe, and P.~Lemieux, ``High
  throughput jpeg 2000 (htj2k): Algorithm, performance and potential,''
  \emph{International Telecommunications Union (ITU)}, pp. 15\,444--15, 2019.

\bibitem{Pranckevicius25}
\BIBentryALTinterwordspacing
A.~Pranckevi{\v{c}}ius, ``Lossless float image compression,'' 07 2025.
  [Online]. Available:
  \url{https://aras-p.info/blog/2025/07/08/Lossless-Float-Image-Compression/}
\BIBentrySTDinterwordspacing

\bibitem{MESHOPTIMIZER}
A.~Kapoulkine, ``meshoptimizer,'' \url{https://github.com/zeux/meshoptimizer},
  2017.

\bibitem{HERSHCOVITCH24}
\BIBentryALTinterwordspacing
M.~Hershcovitch, L.~Choshen, A.~Wood, I.~Enmouri, P.~Chin, S.~Sundararaman, and
  D.~Harnik, ``Lossless and near-lossless compression for foundation models,''
  2024. [Online]. Available: \url{https://arxiv.org/abs/2404.15198}
\BIBentrySTDinterwordspacing

\bibitem{COLLINS18}
M.~Collins, ``Computational graphs, and backpropagation,'' \emph{Lecture Notes,
  Columbia University}, pp. 11--23, 2018.

\bibitem{SHANNON}
C.~Shannon, ``A mathematical theory of communication,'' \emph{The Bell System
  Technical Journal}, vol.~27, pp. 379--423, 1948.

\bibitem{huff}
D.~A. Huffman, ``A method for the construction of minimum-redundancy codes,''
  \emph{Proceedings of the IRE}, vol.~40, no.~9, pp. 1098--1101, 1952.

\bibitem{WITTEN87}
\BIBentryALTinterwordspacing
I.~H. Witten, R.~M. Neal, and J.~G. Cleary, ``Arithmetic coding for data
  compression,'' \emph{Commun. ACM}, vol.~30, no.~6, p. 520–540, Jun. 1987.
  [Online]. Available: \url{https://doi.org/10.1145/214762.214771}
\BIBentrySTDinterwordspacing

\bibitem{duda2014}
\BIBentryALTinterwordspacing
J.~Duda, ``Asymmetric numeral systems: entropy coding combining speed of
  huffman coding with compression rate of arithmetic coding,'' 2014. [Online].
  Available: \url{https://arxiv.org/abs/1311.2540}
\BIBentrySTDinterwordspacing

\bibitem{FSE}
Y.~Collet, ``Finite state entropy - a new breed of entropy coder,''
  \url{https://fastcompression.blogspot.com/2013/12/finite-state-entropy-new-breed-of.html},
  2013.

\bibitem{burrows1994bwt}
\BIBentryALTinterwordspacing
M.~Burrows and D.~J. Wheeler, ``A block-sorting lossless data compression
  algorithm,'' Digital Equipment Corporation, Systems Research Center, Palo
  Alto, CA, SRC Research Report 124, May 1994, sRC Research Report 124.
  [Online]. Available:
  \url{https://www.cs.jhu.edu/~langmea/resources/burrows_wheeler.pdf}
\BIBentrySTDinterwordspacing

\bibitem{MTF}
\BIBentryALTinterwordspacing
J.~L. Bentley, D.~D. Sleator, R.~E. Tarjan, and V.~K. Wei, ``A locally adaptive
  data compression scheme,'' \emph{Commun. ACM}, vol.~29, no.~4, p. 320–330,
  Apr. 1986. [Online]. Available: \url{https://doi.org/10.1145/5684.5688}
\BIBentrySTDinterwordspacing

\bibitem{RLE}
A.~Robinson and C.~Cherry, ``Results of a prototype television bandwidth
  compression scheme,'' \emph{Proceedings of the IEEE}, vol.~55, no.~3, pp.
  356--364, 1967.

\bibitem{lz77}
J.~Ziv and A.~Lempel, ``A universal algorithm for sequential data
  compression,'' \emph{IEEE Transactions on Information Theory}, vol.~23,
  no.~3, pp. 337--343, 1977.

\bibitem{lz4}
Y.~Collet, ``{LZ4 - Extremely fast compression},''
  \url{https://github.com/lz4/lz4}, self-published, Open source project, 2011.

\bibitem{snappy}
\BIBentryALTinterwordspacing
S.~H. Gunderson, ``Snappy: A fast compressor/decompressor,'' 2011, open source
  project. [Online]. Available: \url{https://github.com/google/snappy}
\BIBentrySTDinterwordspacing

\bibitem{rfc1951}
\BIBentryALTinterwordspacing
P.~Deutsch, ``{DEFLATE} compressed data format specification version 1.3,''
  Internet Requests for Comments, RFC Editor, RFC 1951, May 1996, rFC1951.
  [Online]. Available: \url{https://www.rfc-editor.org/rfc/rfc1951.txt}
\BIBentrySTDinterwordspacing

\bibitem{rfc1952}
\BIBentryALTinterwordspacing
L.~P. Deutsch, ``{GZIP file format specification version 4.3},'' RFC 1952, May
  1996. [Online]. Available:
  \url{https://datatracker.ietf.org/doc/html/rfc1952}
\BIBentrySTDinterwordspacing

\bibitem{rfc7932}
\BIBentryALTinterwordspacing
J.~Alakuijala and Z.~Szabadka, ``Brotli compressed data format,'' Internet
  Requests for Comments, RFC Editor, RFC 7932, July 2016. [Online]. Available:
  \url{https://www.rfc-editor.org/rfc/rfc7932.txt}
\BIBentrySTDinterwordspacing

\bibitem{rfc8878}
\BIBentryALTinterwordspacing
Y.~Collet and M.~Kucherawy, ``{Zstandard} compression and the application/zstd
  media type,'' Internet Requests for Comments, RFC Editor, RFC 8878, October
  2018, rFC8878. [Online]. Available:
  \url{https://www.rfc-editor.org/rfc/rfc8878.txt}
\BIBentrySTDinterwordspacing

\bibitem{LZMA}
\BIBentryALTinterwordspacing
I.~Pavlov, ``{LZMA} algorithm description,'' 2013, 7-zip documentation.
  [Online]. Available: \url{https://www.7-zip.org/7z.html}
\BIBentrySTDinterwordspacing

\bibitem{ppm}
J.~Cleary and I.~Witten, ``Data compression using adaptive coding and partial
  string matching,'' \emph{IEEE Transactions on Communications}, vol.~32,
  no.~4, pp. 396--402, 1984.

\bibitem{dmc}
\BIBentryALTinterwordspacing
G.~V. Cormack and R.~N.~S. Horspool, ``Data compression using dynamic markov
  modelling,'' \emph{The Computer Journal}, vol.~30, no.~6, pp. 541--550, 12
  1987. [Online]. Available: \url{https://doi.org/10.1093/comjnl/30.6.541}
\BIBentrySTDinterwordspacing

\bibitem{mahoney2013paq}
\BIBentryALTinterwordspacing
M.~Mahoney, ``The {PAQ} data compression programs,'' 2013, website documenting
  various PAQ iterations. [Online]. Available:
  \url{http://mattmahoney.net/dc/paq.html}
\BIBentrySTDinterwordspacing

\bibitem{rfc2083}
\BIBentryALTinterwordspacing
T.~Boutell, ``{PNG (Portable Network Graphics) Specification Version 1.0},''
  RFC 2083, March 1997. [Online]. Available:
  \url{https://www.rfc-editor.org/info/rfc2083}
\BIBentrySTDinterwordspacing

\bibitem{BLOSC}
\BIBentryALTinterwordspacing
T.~B.~D. Team, ``Blosc documentation,'' 2010. [Online]. Available:
  \url{https://www.blosc.org}
\BIBentrySTDinterwordspacing

\bibitem{PARQUET}
A.~S. Foundation, ``parquet-format,''
  \url{https://parquet.apache.org/docs/file-format/data-pages/compression/},
  Apache Software Foundation, Tech. Rep., 2024.

\bibitem{BTUNE}
T.~B.~D. Team, ``Btune: Making compression better,''
  \url{https://blosc.org/pages/btune}, IronArray SLU, Tech. Rep., 2023.

\bibitem{mahoney2015zpaq}
\BIBentryALTinterwordspacing
M.~Mahoney, ``The zpaq compression algorithm,'' self-published, Technical
  Report ZPAQ-2015-12-29, Dec. 2015. [Online]. Available:
  \url{https://mattmahoney.net/dc/zpaq_compression.pdf}
\BIBentrySTDinterwordspacing

\bibitem{DYER16}
\BIBentryALTinterwordspacing
C.~Dyer, Y.~Goldberg, and G.~Neubig, ``Practical neural networks for {NLP}:
  From theory to code,'' in \emph{Proceedings of the 2016 Conference on
  Empirical Methods in Natural Language Processing: Tutorial Abstracts},
  B.~Yang and R.~Hwa, Eds.\hskip 1em plus 0.5em minus 0.4em\relax Austin,
  Texas: Association for Computational Linguistics, Nov. 2016. [Online].
  Available: \url{https://aclanthology.org/D16-2001/}
\BIBentrySTDinterwordspacing

\bibitem{SILESIA}
S.~Deorowicz, ``Universal lossless data compression algorithms,'' Ph.D.
  dissertation, Silesian University of Technology, 2003.

\bibitem{SAO}
\BIBentryALTinterwordspacing
``Sao star catalog,'' 2002. [Online]. Available:
  \url{http://tdc-www.harvard.edu/software/catalogs/sao.html}
\BIBentrySTDinterwordspacing

\bibitem{BINANCE}
J.~Smit, ``Binance full history,''
  \url{https://www.kaggle.com/datasets/jorijnsmit/binance-full-history}, 2025.

\bibitem{TLC}
``{TLC} trip record data,''
  \url{https://www.nyc.gov/site/tlc/about/tlc-trip-record-data.page}, 2022.

\bibitem{ERA5}
H.~Hersbach, B.~Bell, P.~Berrisford, G.~Biavati, A.~Horányi,
  J.~Muñoz~Sabater, J.~Nicolas, C.~Peubey, R.~Radu, I.~Rozum, D.~Schepers,
  A.~Simmons, C.~Soci, D.~Dee, and J.-N. Thépaut, ``Era5 hourly data on single
  levels from 1940 to present,'' 2023.

\bibitem{ppmf2024}
\BIBentryALTinterwordspacing
U.~C. Bureau, ``2020 census privacy-protected microdata file (ppmf) readme,''
  2024. [Online]. Available:
  \url{https://www2.census.gov/programs-surveys/decennial/2020/data/privacy-protected-microdata-file/2024-08-05-privacy-protected-microdata-file-README.pdf}
\BIBentrySTDinterwordspacing

\bibitem{pums2023}
\BIBentryALTinterwordspacing
A.~C.~S. Office, ``5-year pums data (2023),'' 2024. [Online]. Available:
  \url{https://www2.census.gov/programs-surveys/acs/data/pums/2023/5-Year/}
\BIBentrySTDinterwordspacing

\bibitem{DEB02}
K.~Deb, A.~Pratap, S.~Agarwal, and T.~Meyarivan, ``A fast and elitist
  multiobjective genetic algorithm: Nsga-ii,'' \emph{IEEE Transactions on
  Evolutionary Computation}, vol.~6, no.~2, pp. 182--197, 2002.

\bibitem{koza92}
J.~R. Koza, \emph{Genetic Programming: On the Programming of Computers by Means
  of Natural Selection}.\hskip 1em plus 0.5em minus 0.4em\relax Cambridge, MA,
  USA: MIT Press, 1992.

\bibitem{JEONG22}
G.~Jeong, B.~Sharma, N.~Terrell, A.~Dhanotia, Z.~Zhao, N.~Agarwal,
  A.~Kejariwal, and T.~Krishna, ``Understanding data compression in
  warehouse-scale datacenter services,'' in \emph{2022 IEEE International
  Symposium on Performance Analysis of Systems and Software (ISPASS)}, 2022,
  pp. 221--223.

\bibitem{thrift}
\BIBentryALTinterwordspacing
M.~Slee, A.~Agarwal, and M.~Kwiatkowski, ``Thrift: Scalable cross-language
  services implementation,'' Facebook, Technical Report, 2007. [Online].
  Available: \url{https://thrift.apache.org/static/files/thrift-20070401.pdf}
\BIBentrySTDinterwordspacing

\bibitem{NIMBLE}
(2024) nimble. \url{https://github.com/facebookincubator/nimble}. Facebook
  Incubator.

\bibitem{SCRIBE}
M.~Karpathiotakis, D.~Wernli, and M.~Stojanovic. (2019) Scribe: Transporting
  petabytes per hour via a distributed, buffered queueing system.
  \url{https://engineering.fb.com/2019/10/07/core-infra/scribe/}.

\bibitem{HANDTE18}
\BIBentryALTinterwordspacing
W.~F. Handte, Y.~Collet, and N.~Terrell, ``5 ways {Facebook} improved
  compression at scale with {Zstandard},'' 2018. [Online]. Available:
  \url{https://engineering.fb.com/2018/12/19/core-infra/zstandard/}
\BIBentrySTDinterwordspacing

\end{thebibliography}


\end{document}